\documentclass[twocolumn]{aastex61}

\usepackage[utf8x]{inputenc}

\usepackage{graphicx}
\usepackage{float}
\usepackage{multirow}

\begin{document}

\title{ARES\footnote{Ariel Retrieval Exoplanet School} V: No Evidence For Molecular Absorption in the HST WFC3 Spectrum of GJ 1132 b}

\correspondingauthor{Lorenzo V. Mugnai}
\email{lorenzo.mugnai@uniroma1.it}

\author[0000-0002-9007-9802]{Lorenzo V. Mugnai} 
\affil{La Sapienza Universit\'a di Roma, Department of Physics, Piazzale Aldo Moro 2, 00185 Roma, Italy}

\author[0000-0001-6425-9415]{Darius Modirrousta-Galian} 
\affil{INAF – Osservatorio Astronomico di Palermo, Piazza del Parlamento 1, I-90134 Palermo, Italy}
\affil{University of Palermo, Department of Physics and Chemistry, Via Archirafi 36, Palermo, Italy}

\author[0000-0002-5494-3237]{Billy Edwards}
\affil{Department of Physics and Astronomy, University College London, London, United Kingdom}

\author[0000-0001-6516-4493]{Quentin Changeat}
\affil{Department of Physics and Astronomy, University College London, London, United Kingdom}

\author{Jeroen Bouwman} 
\affil{Max-Planck-Institut für Astronomie, Königstuhl 17, 69117 Heidelberg, Germany}

\author[0000-0002-4262-5661]{Giuseppe Morello} 
\affil{Instituto de Astrof\'isica de Canarias (IAC), 38205 La Laguna, Tenerife, Spain}
\affil{INAF – Osservatorio Astronomico di Palermo, Piazza del Parlamento 1, I-90134 Palermo, Italy}

\author[0000-0003-2241-5330]{Ahmed Al-Refaie} 
\affil{Department of Physics and Astronomy, University College London, London, United Kingdom}

\author{Robin Baeyens}
\affil{Institute of Astronomy, KU~Leuven, Celestijnenlaan 200D, 3001 Leuven, Belgium}

\author{Michelle Fabienne Bieger}
\affil{College of Engineering, Mathematics and Physical Sciences, Physics Building, University of Exeter, Exeter, United Kingdom}

\author{Doriann Blain}
\affil{LESIA, Observatoire de Paris, Université PSL, CNRS, Sorbonne Universit\'e, Universit\'e de Paris, Meudon, France}

\author{Am\'elie Gressier}
\affil{Sorbonne Universit\'es, UPMC Universit\'e Paris 6 et CNRS, 
UMR 7095, Institut d'Astrophysique de Paris, Paris, France}
\affil{LESIA, Observatoire de Paris, Université PSL, CNRS, Sorbonne Universit\'e, Universit\'e de Paris, Meudon, France}

\author{Gloria Guilluy}
\affil{Dipartimento di Fisica, Universit\'a degli Studi di Torino, via Pietro Giuria 1, I-10125 Torino, Italy}
\affil{INAF Osservatorio Astrofisico di Torino, Via Osservatorio 20, I-10025 Pino Torinese, Italy}

\author{Yassin Jaziri}  
\affil{Laboratoire d'astrophysique de Bordeaux, Univ. Bordeaux, CNRS, B18N, all\'{e}e Geoffroy Saint-Hilaire, 33615 Pessac, France}

\author{Flavien Kiefer}
\affil{LESIA, Observatoire de Paris, Université PSL, CNRS, Sorbonne Universit\'e, Universit\'e de Paris, Meudon, France}

\author{Mario Morvan}
\affil{Department of Physics and Astronomy, University College London, London, United Kingdom}

\author{William Pluriel}  
\affil{Laboratoire d'astrophysique de Bordeaux, Univ. Bordeaux, CNRS, B18N, all\'{e}e Geoffroy Saint-Hilaire, 33615 Pessac, France}

\author{Mathilde Poveda}
\affil{Laboratoire Interuniversitaire des Systèmes Atmosphériques (LISA), UMR CNRS 7583, Universit\'e Paris-Est-Cr\'eteil, Universit\'e de Paris, Institut Pierre Simon Laplace, Créteil, France}
\affil{Maison de la Simulation, CEA, CNRS, Univ. Paris-Sud, UVSQ, Universit\'e Paris-Saclay, F-91191 Gif-sur-Yvette, France}

\author[0000-0002-9372-5056]{Nour Skaf} 
\affil{Department of Physics and Astronomy, University College London, London, United Kingdom}
\affil{LESIA, Observatoire de Paris, Université PSL, CNRS, Sorbonne Universit\'e, Universit\'e de Paris, Meudon, France}
\affil{Subaru Telescope, National Astronomical Observatory of Japan, 650 North A'ohoku Place, Hilo, HI 96720, USA}

\author{Niall Whiteford}
\affil{Institute for Astronomy, University of Edinburgh, Royal Observatory, Blackford Hill, Edinburgh, EH9 3HJ, UK}
\affil{Centre for Exoplanet Science, University of Edinburgh, UK}

\author{Sam Wright}
\affil{Department of Physics and Astronomy, University College London, London, United Kingdom}

\author{Kai Hou Yip}
\affil{Department of Physics and Astronomy, University College London, London, United Kingdom}

\author{Tiziano Zingales} 
\affil{Laboratoire d'astrophysique de Bordeaux, Univ. Bordeaux, CNRS, B18N, all\'{e}e Geoffroy Saint-Hilaire, 33615 Pessac, France}

\author{Benjamin Charnay}
\affil{LESIA, Observatoire de Paris, Université PSL, CNRS, Sorbonne Universit\'e, Universit\'e de Paris, Meudon, France}

\author{Pierre Drossart}  
\affil{Sorbonne Universit\'es, UPMC Universit\'e Paris 6 et CNRS, 
UMR 7095, Institut d'Astrophysique de Paris, Paris, France}
\affil{LESIA, Observatoire de Paris, Université PSL, CNRS, Sorbonne Universit\'e, Universit\'e de Paris, Meudon, France}

\author{J\'{e}r\'{e}my Leconte}  
\affil{Laboratoire d'astrophysique de Bordeaux, Univ. Bordeaux, CNRS, B18N, all\'{e}e Geoffroy Saint-Hilaire, 33615 Pessac, France}

\author{Olivia Venot}  
\affil{Laboratoire Interuniversitaire des Systèmes Atmosphériques (LISA), UMR CNRS 7583, Universit\'e Paris-Est-Cr\'eteil, Universit\'e de Paris, Institut Pierre Simon Laplace, Créteil, France}

\author[0000-0002-4205-5267]{Ingo Waldmann}
\affil{Department of Physics and Astronomy, University College London, London, United Kingdom}

\author{Jean-Philippe Beaulieu}
\affil{School of Physical Sciences, University of Tasmania,
Private Bag 37 Hobart, Tasmania 7001 Australia}
\affil{Sorbonne Universit\'es, UPMC Universit\'e Paris 6 et CNRS, 
UMR 7095, Institut d'Astrophysique de Paris, Paris, France}

%%ABSTRACT

\begin{abstract}

We present a study on the spatially scanned spectroscopic observations of the transit of GJ~1132~b, a warm ($\sim$500 K) Super-Earth (1.13 R$_\oplus$) that was obtained with the G141 grism (1.125 - 1.650 $\mu$m) of the Wide Field Camera 3 (WFC3) onboard the Hubble Space Telescope. We used the publicly available Iraclis pipeline to extract the planetary transmission spectra from the five visits and produce a precise transmission spectrum. We analysed the spectrum using the TauREx3 atmospheric retrieval code with which we show that the measurements do not contain molecular signatures in the investigated wavelength range and are best-fit with a flat-line model. Our results suggest that the planet does not have a clear primordial, hydrogen-dominated atmosphere. Instead, GJ~1132~b could have a cloudy hydrogen-dominated envelope, a very enriched secondary atmosphere, be airless, or have a tenuous atmosphere that has not been detected. Due to the narrow wavelength coverage of WFC3, these scenarios cannot be distinguished yet but the James Webb Space Telescope may be capable of detecting atmospheric features, although several observations may be required to provide useful constraints. \vspace{10mm}

\end{abstract}

%%KEYWORDS
%\keywords{methods: data analysis; methods: statistical; planets and satellites: atmospheres; Astrophysics - Earth and Planetary Astrophysics }

% _____________________________________________
% _____________________________________________

\section{Introduction}

One major obstacle that exoplanetary researchers encounter is a general lack of data. This makes it difficult to determine the composition and internal structure of exoplanets as there is an inevitable strong degeneracy when one tries to fit a model to observations. By making use of geophysical and statistical principles, several studies have determined the degree of degeneracy in exoplanet compositions \citep[e.g.][]{Adams2008,Valencia2013,Dorn2017}. They found that knowing the mass and radius of a planet precisely can lead to superior constraints on the ice mass fraction and size of the inner embryo, but little improvement on the atmospheric composition. However, they also found that determining the atmospheric composition (such as from spectroscopy) could lead to a significant improvement of the interior predictions. Therefore, there is a strong motivation to characterise exoplanetary atmospheres as this would lead to a better understanding on the global properties of their host planets.

In spite of this, under most circumstances only the mass and radius of exoplanets are known, so all that can be done is constrain the internal compositions from the bulk mean densities \citep[e.g.][]{Zeng2013,Zeng2016}. Recent advances in exoplanetary spectroscopy have allowed for the atmospheric composition and structure to be constrained enough to attain a more holistic understanding of the planet. For instance, from the mass and radius of a planet one cannot tell whether a super-Earth or sub-Neptune is $\rm H_{2}O$-rich or a silicate embryo with a hydrogen envelope \citep[e.g.][]{Valencia2013}. However, if one were to constrain the atmospheric composition of these perplexing bodies, then one could determine whether the planet is icy (i.e. no atmosphere or a $\rm H_{2}O$-rich one if the temperature is high enough) or rocky with a hydrogen-rich atmosphere (i.e. collisional absorption lines of hydrogen are detected with, perhaps, some mineral or volcanic species).

A number of studies using HST WFC3 have found evidence for molecular absorption in Sub-Neptunes \citep[e.g.][]{Guo2020,ares4}. Of particular note are the studies of the habitable-zone planet K2-18\,b, which likely has a hydrogen–helium envelope with a high concentration of water vapor \citep{Tsiaras2019,Benneke2019} and possibly $\rm CH_{4}$ \citep{Bezard2020,Blain2020}. Meanwhile, GJ~1214~b most probably hosts a thick cloud layer, with molecular features belonging to cloud-free primary atmosphere, or one composed of $100\%$ $\rm H_{2}O$ and $100\%$ $\rm CO_{2}$, having been ruled out \citep[e.g.][]{Kreidberg2014}.

Whilst the atmospheric spectroscopy of small, potentially rocky, exoplanets is difficult, several analyses have already been made on well-known systems. For example:
\begin{itemize}
    \item TRAPPIST-1 b, c, d, e, f and g most probably do not have cloud-free hydrogen atmospheres \citep[e.g.][]{deWit2016}.
    \item The HST WFC3 transmission spectrum of the highly irradiated Super-Earth 55 Cnc e shows evidence for Hydrogen Cyanide (HCN) \citep{Tsiaras2016a}. However, the exact nature of its atmosphere, and whether it exists, is still highly debated \citep{Madhusudhan2012,Dorn2019,Modirrousta2020a,Jindal2020,Zilinskas2021,Zhang2020}
    \item LHS~1140~b, a super-Earth orbiting in the habitable-zone of its star, potentially hosts an atmosphere containing water vapour but the low SNR, and narrow wavelength coverage, of the data means this detection is tentative \citep{edwards_lhs}.
\end{itemize}

Additionally, the Spitzer phase curve of the terrestrial planet LHS~3844~b is incompatible with a thick atmosphere \citep{Kreidberg2019}. Thus, so far, there have been no definitive measurements of the atmosphere of a rocky exoplanet.

\begin{figure}
  		\centering
   		\includegraphics[width = \columnwidth]{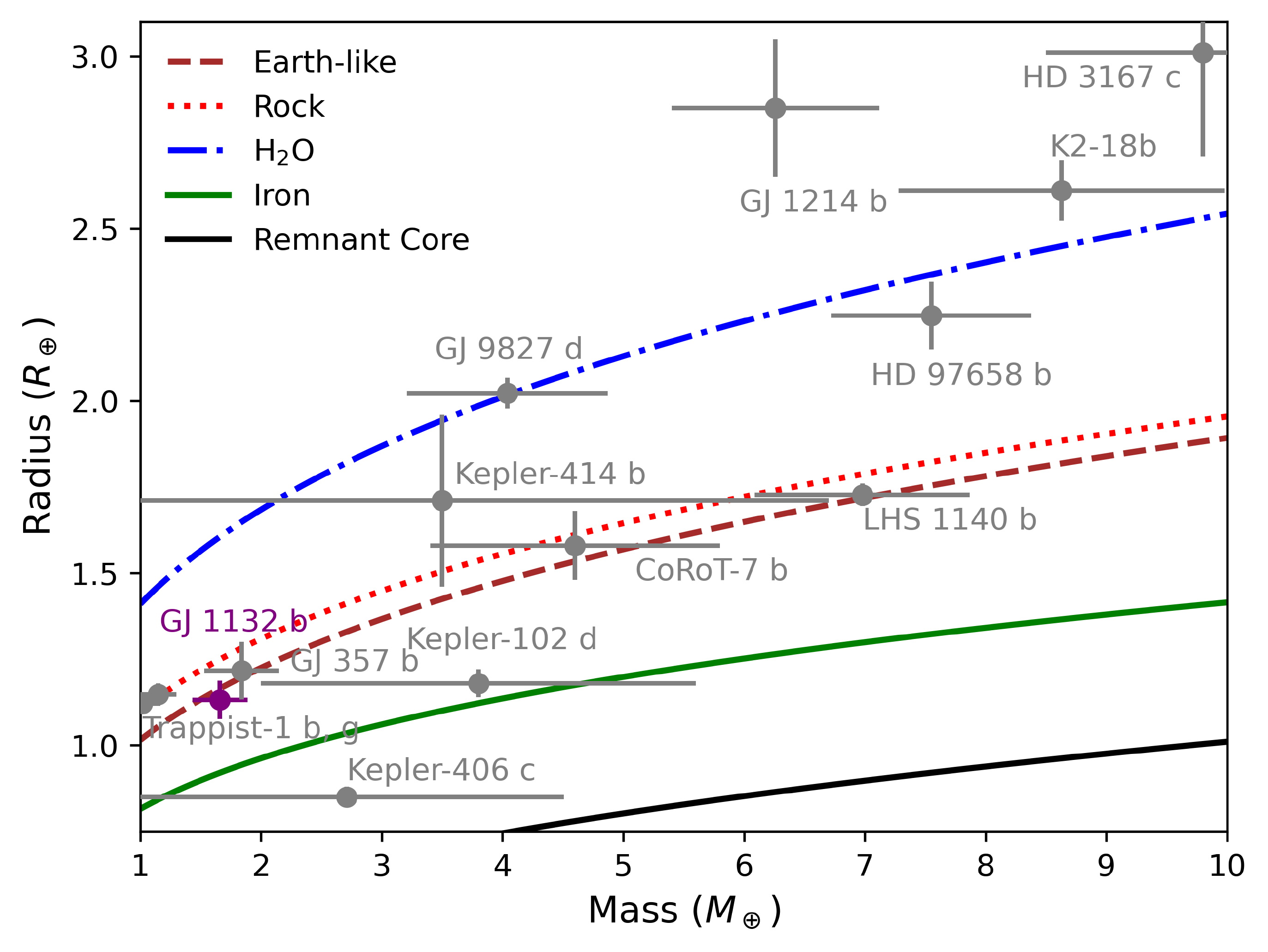}
   		\caption{The mass and radius plot of GJ~1132~b \citep{Bonfils2018} with other super-Earths and sub-Neptunes. The Earth-like, $\rm H_{2}O$ and Iron mass and radius models are from \citet{Zeng2013} and \citet{Zeng2016}. The remnant core model (i.e. a planet that is highly compressed due to once hosting a large primordial atmosphere that was then subsequently lost) is from \citet{Mocquet2014}. It is important to note that the mass-radius models shown above are merely illustrative as more complex set-ups are possible \citep[e.g.][]{Jespersen2020,Mousis2020,Modirrousta2020a}. The planets listed are Trappist-1~b,g \citep{Grimm2018,deWit2016}, GJ~357~b \citep{Luque2019}, Kepler-406~c \citep{Marcy2014}, Kepler-414~b \citep{Hadden2014}, Kepler-102~d \citep{Marcy2014}, GJ~9827~d \citep{Rice2019}, GJ~1214~b \citep{Harpsoe2013},CoRoT-7~b \citep{Dai2019}, HD~97658~b \citep{VanGrootel2014,Guo2020}, HD~3167~c \citep{Christiansen2017,ares4,Mikal2021}, LHS~1140~b \citep{Ment2019}, and K2-18b \citep{Benneke2019,Tsiaras2019,Bezard2020,Blain2020}}.
   		\label{fig:mass_radius}
   		\centering
   	\end{figure}

In this paper we perform a spectroscopic analysis of GJ\,1132\,b with the aim of determining its atmospheric composition. Making use of the mass and radius measurements from the literature, we then make inferences on the interior composition and properties of GJ\,1132\,b. Having a mass, radius, and equilibrium temperature of $\rm 1.66 \pm 0.23\,M_{\oplus}$, $\rm 1.130 \pm 0.056\,R_{\oplus}$, and $\rm 500-600~K$ \citep{Bonfils2018} respectively, GJ\,1132\,b is a super-Earth that may be an ice planet that migrated inwards, or a silicate embryo with a hydrogen envelope (see Figure~\ref{fig:mass_radius}). A mixture of these two compositions or a more exotic make-up may also be possible \citep{Zeng2016}. 

In order to try to overcome this degeneracy we perform a spectroscopic analysis on the spectral data obtained through \textit{Hubble} observations. Using five transit observations, we recover a flat spectrum which shows no sign of atmospheric features. We rule out a clear hydrogen/helium dominated atmosphere to $>5 \sigma$. Future observations are required to confidently distinguish between a cloudy primary atmosphere or one with a higher mean molecular weight.

% _____________________________________________
% _____________________________________________

\section{Method}

\subsection{Data analysis}
Our analysis is based on five transit observations of GJ\,1132\,b (Table \ref{tab:iraclis_input}) obtained between April and November 2017 with the G141 infrared grism ($1.125 - 1.650 \; \mu m $) of the Wide Field Camera 3 (WFC3) on board the Hubble Space Telescope (HST). The observations were part of the HST proposal number 14758 led by Zach Berta-Thompson \citep{berta_prop}, and were downloaded from the public Mikulski Archive for Space Telescope (MAST) archive\footnote{\url{https://archive.stsci.edu/hst/}}. 

Each transit observation required 4 HST orbits and utilised the spatial scanning technique. The observations were acquired using the 256x256 sub-array, employing the SPARS10 sampling sequence with 15 up-the-ramp reads which lead to an exposure time of 103.129 s. The scan speed was 0.2"/s leading to a total scan length of 170 pixels and a maximum pixel fluence below 24,000 electrons.

\begin{table*}[]
    \centering
    \caption{Star and planet parameters used as reference in this paper.}
    \label{tab:iraclis_input}
    \begin{tabular}{c c c}
         \hline
         \hline
         Parameter &  Value & Source \\
         \hline
         \multicolumn{3}{c}{Stellar parameters} \\
         \hline
         Spectral type & M4.5 V & \citet{Berta-Thompson2015} \\
       $T_{\rm eff} \; [K]$ & $3270 \pm 140 $ & \citet{Bonfils2018} \\
       $R_\star \; [R_\sun]$ & $0.2105 ^{+0.0102}_{-0.0085}$& \citet{Bonfils2018}\\
       $\log_{10} g \; [cm/s^2]$ & $5.05 \pm 0.074$ & \citet{Bonfils2018} \\
        $[Fe/H]$ & $-0.12 \pm 0.15$ & \citet{Berta-Thompson2015} \\
          \hline
      \multicolumn{3}{c}{Planetary parameters} \\
      \hline
      $M_p \; [M_\Earth]$ & $1.66 \pm 0.23$ & \citet{Bonfils2018} \\
      $R_p \; [R_\Earth]$ & $1.130 \pm 0.056$ &
      \citet{Bonfils2018} \\
      $P \; [day]$ & $1.628931 \pm 0.000027$ & \citet{Bonfils2018}\\
      $a/R_\star$ & $15.6235^{+0.8}_{-0.7}$ & \citet{Bonfils2018}$^a$\\
      $e$ & $<0.22^b$ & \citet{Bonfils2018}\\
      $i \; [deg]$ & $88.68 ^{+0.40}_{-0.33}$ & \citet{Bonfils2018}\\
      $T_{\text{mid}} \; [BJD - 2450000]$ & $7184.55786 \pm 0.00031$ & \citet{Bonfils2018}\\\hline
      %$(R_p /R_\star)^2$ & $0.00262 \pm 0.00026$ & \citet{Berta-Thompson2015}
      \multicolumn{3}{c}{$^a$derived from \citep{Bonfils2018} measurements of $a$ and $R_\star$}\\
       \multicolumn{3}{c}{$^b$ fixed to zero as in \citet{Berta-Thompson2015}}\\
         \hline  \hline
    \end{tabular}
\end{table*}

Following the procedure already described in similar studies \citep[e.g.][]{ares1, ares2, anisman_w117,kelt9}, we extract white and spectral light-curves from the raw HST/WFC3 images using the Iraclis software \citep{Tsiaras2016b}. Iraclis is an open-source software dedicated to the analysis of WFC3 scanning observations and is publicly available on GitHub\footnote{\url{https://github.com/ucl-exoplanets/Iraclis}}. In the following we briefly summarise the data analysis steps operated by the software, but we refer the reader to \citet{Tsiaras2016b} for a complete discussion.

\subsubsection{Data reduction and calibration}
\begin{figure*}
    \centering
    \includegraphics[width=0.95\textwidth]{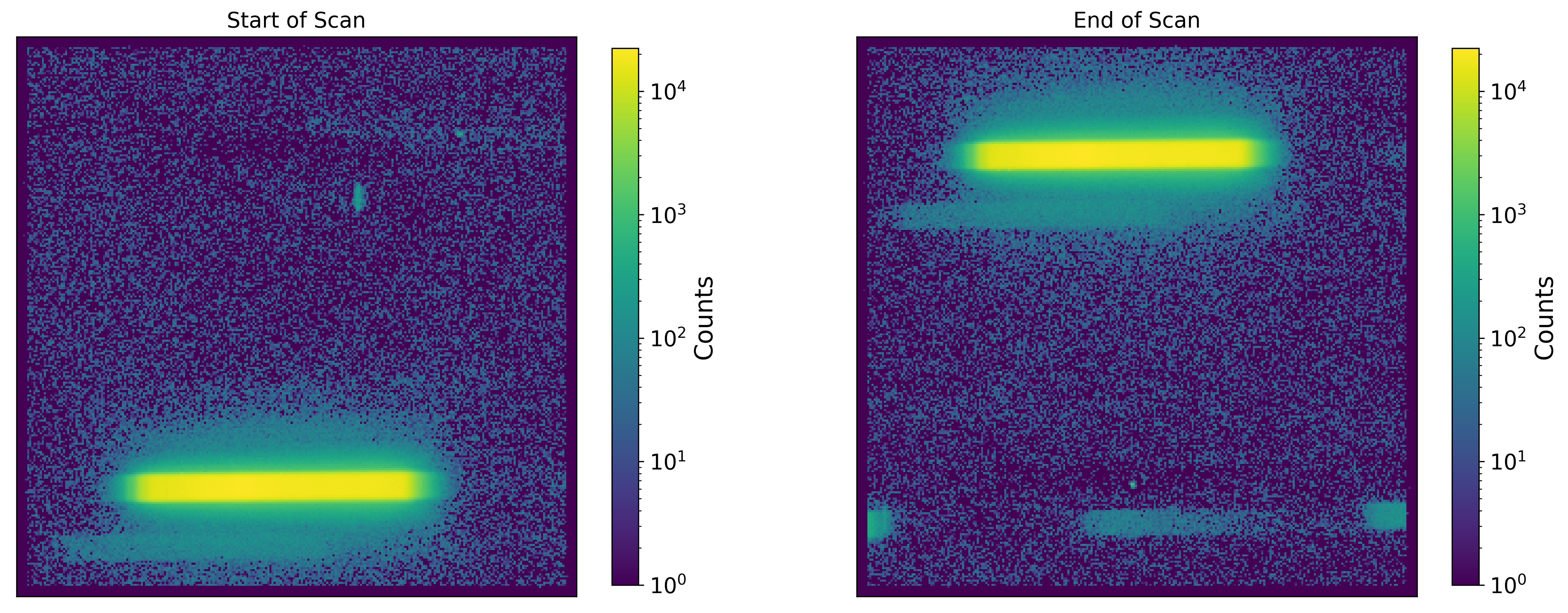}
    \caption{Example reads from the beginning (left) and end (right) of a scan for the second visit. If the data were not extracted from each read individually, the two faint background sources would contribute to the spectrum. The colour indicates the count per pixel and is represented with a log scale.}
    \label{fig:split_data}
\end{figure*}

The Iraclis reduction process included the following steps: zero-read subtraction, reference pixels correction, non-linearity correction, dark current subtraction, gain conversion, sky background subtraction, calibration, flat-field correction, and corrections for bad pixels and cosmic rays \citep{Tsiaras2016b}. In each of the visits, we noted several faint sources which overlapped with the spectrum for GJ\,1132. Hence we split these data into individual up-the-ramp reads and performed the extraction on these to remove the signal from these secondary sources which could impact the shorter wavelengths of the recovered spectrum. Two example reads from the second visit are shown in Figure \ref{fig:split_data}, with the first order spectrum from faint source visible below the data from GJ\,1132 and the zeroth order of another star also within the scan's arc.

\subsubsection{Light curve extraction}

We extract the wavelength-dependent light-curves taking into account of the geometric distortions caused by the tilted detector of the WFC3/IR channel.

We obtain a \textit{white light curve}, and a set of \textit{spectral light curves}. The former is obtained integrating the full wavelength range of WFC3/G141, while the latter are extracted using narrow band such that the resolving power is 70 at $1.4 \; \mu m$.

\subsubsection{Limb darkening coefficients}
The limb-darkening coefficients are computed using the non-linear formula by \citet{Claret2000}, which scales the intensity emerging from the star as:
\begin{equation}
    \frac{I(\mu)}{I(1)} = 1 - \sum_{k=1}^4 a_k(1-\mu^\frac{k}{2})
\end{equation}
where $\mu= \cos(\gamma)$, and $\gamma$ is the angle between the line of sight and the emergent intensity, and $a_k$ are the limb darkening coefficients.

We calculated the $a_k$ coefficients using \texttt{ExoTETHyS} \citep{Morello2020}, with the stellar models from Phoenix 2018 \citep{claret_2018} and the parameters in Table \ref{tab:iraclis_input}. These are given in Table \ref{tab:limb_darkening}.

\begin{table}[]
    \centering
    \caption{Limb darkening coefficients used here. The first row represents the white light curve while all others are for the spectral light curves.}
    \begin{tabular}{cccccc}
    \hline\hline
    $ \lambda \; [\mu m]$ & a1 & a2 & a3 & a4 & $ \Delta \lambda \; [\mu m]$  \\
    \hline
    1.3840 & 1.475 & -1.351 & 0.830 & 0.592 \\\hline
1.1262 & 1.461 & -1.439 & 0.922 & -0.247 & 0.0219\\
1.1478 & 1.435 & -1.431 & 0.924 & -0.249 & 0.0211\\
1.1686 & 1.413 & -1.412 & 0.912 & -0.246 & 0.0206\\
1.1888 & 1.430 & -1.443 & 0.934 & -0.252 & 0.0198\\
1.2084 & 1.406 & -1.406 & 0.907 & -0.244 & 0.0193\\
1.2275 & 1.363 & -1.363 & 0.880 & -0.237 & 0.0190\\
1.2465 & 1.364 & -1.361 & 0.876 & -0.236 & 0.0189\\
1.2655 & 1.361 & -1.371 & 0.886 & -0.239 & 0.0192\\
1.2848 & 1.324 & -1.330 & 0.859 & -0.232 & 0.0193\\
1.3038 & 1.318 & -1.333 & 0.862 & -0.233 & 0.0188\\
1.3226 & 1.361 & -1.386 & 0.896 & -0.241 & 0.0188\\
1.3415 & 1.376 & -1.247 & 0.759 & -0.198 & 0.0189\\
1.3605 & 1.447 & -1.314 & 0.787 & -0.202 & 0.0192\\
1.3800 & 1.526 & -1.468 & 0.904 & -0.236 & 0.0199\\
1.4000 & 1.477 & -1.337 & 0.799 & -0.205 & 0.0200\\
1.4202 & 1.456 & -1.275 & 0.746 & -0.190 & 0.0203\\
1.4406 & 1.475 & -1.288 & 0.745 & -0.187 & 0.0206\\
1.4615 & 1.417 & -1.183 & 0.672 & -0.168 & 0.0212\\
1.4831 & 1.461 & -1.273 & 0.735 & -0.184 & 0.0220\\
1.5053 & 1.475 & -1.316 & 0.766 & -0.193 & 0.0224\\
1.5280 & 1.475 & -1.332 & 0.784 & -0.199 & 0.0230\\
1.5515 & 1.452 & -1.309 & 0.772 & -0.197 & 0.0241\\
1.5762 & 1.478 & -1.374 & 0.816 & -0.208 & 0.0253\\
1.6021 & 1.485 & -1.435 & 0.872 & -0.225 & 0.0264\\
1.6295 & 1.475 & -1.476 & 0.917 & -0.240 & 0.0283\\
\hline
    \end{tabular}
    \label{tab:limb_darkening}
\end{table}

%We chose the $a_k$ coefficients by fitting a stellar profile from an ATLAS model \citep{Kurucz1970, Howarth2011, Morello2020} referring to Table \ref{tab:iraclis_input} for the stellar parameters. 

\subsubsection{White light-curves fitting}
\label{sec:white_curve}

To fit the extracted white and spectral light-curves, we consider the known time-dependent systematics introduced by HST: 
\begin{itemize}
    \item long-term `ramp', characterised by a linear trend
    \item short-term `ramp', characterised by an exponential trend.
\end{itemize}
To remove these systematics, we multiply for a normalisation factor $n_w$ and an instrumental corrective factor $R(t)$. The former depend on the telescope observing scanning mode, $n_w^{\text{scan}}$, and when scanning direction is upwards changes to $n_{w}^{for}$ or to $n_{w}^{\text{rev}}$ when scanning direction is downwards. $R(t)$ is time dependent and can be derived as 
\begin{equation}
R(t)  = (1-r_a(t-T_0))\left( 1-r_{b_1} e^{-r_{b_2}(t-t_0)} \right)
   \label{eq1}
\end{equation}
where $t$ is time, $T_0$ is the mid-transit time, $t_0$ is the starting time of each HST orbit, $r_{a}$ is the linear systematic trend's slope, $r_{b_1}$ and $r_{b_2}$ are the exponential systematic trend's coefficients \citep{ Kreidberg2014,Tsiaras2016a, Tsiaras2016b} .

To extract the white transit light curve, $F_w^v(t)$, for each visit, $v$, we both fit the transit model multiplied by the instrumental systematics, $n_w^{\text{scan}} R(t) F_w^v(t)$ and we fit the systematic model $R(t)$ on the out of transit data, correcting the light curve and then fitting again $n_{w}^{\text{scan}} F_w^v(t)$.

\begin{figure*}
    \centering
    \includegraphics[width = \textwidth]{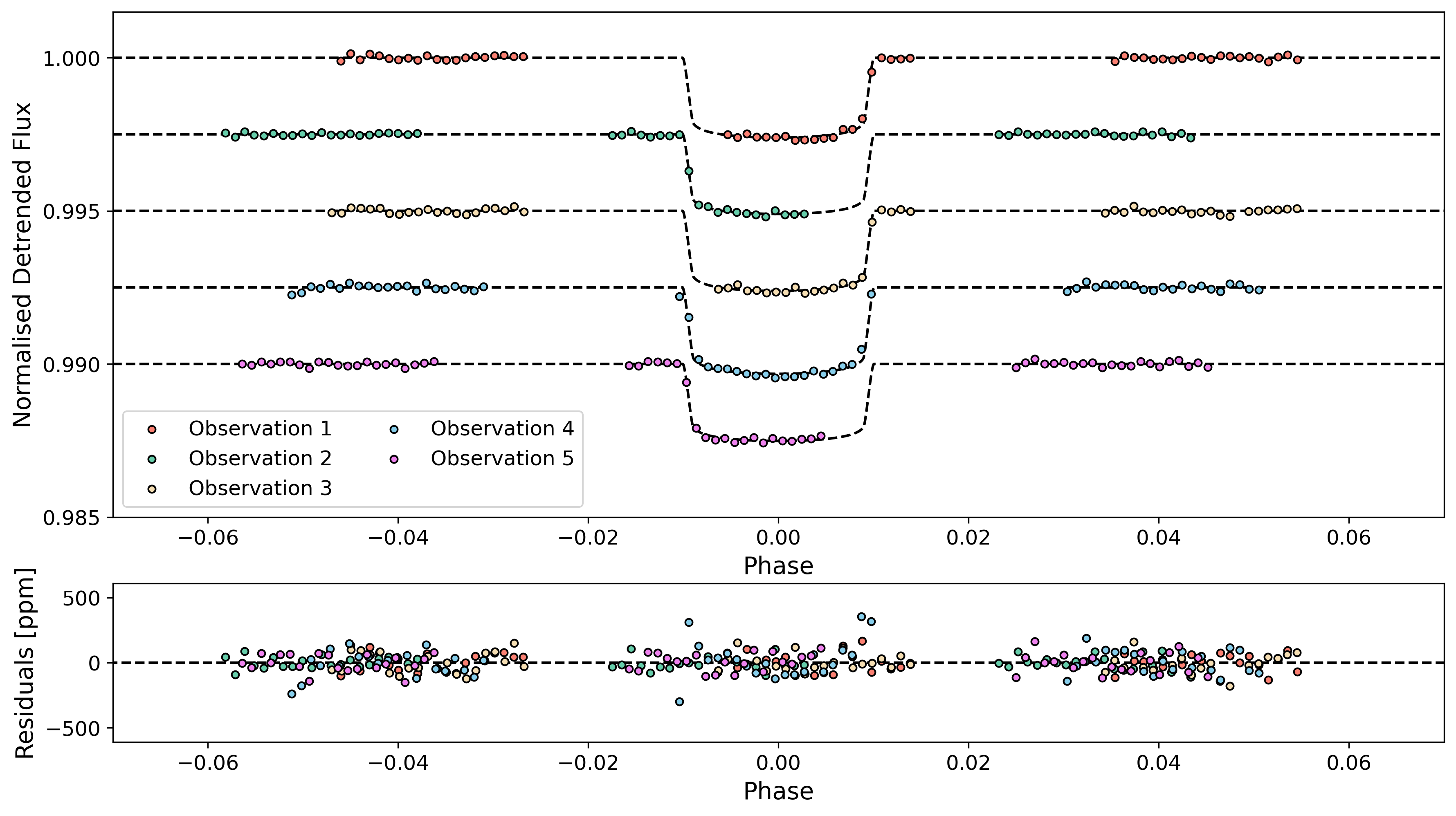}
    \caption{In the top panel are reported the detrended white light curves from each visit along with the best-fit transit model. In the bottom panel, the residuals are reported. \label{fig:white_lc} }
\end{figure*}

To fit the light curves we use the data presented in Table \ref{tab:iraclis_input}, leaving as free parameters the radii ratio, $R_p/R_\star$, and the mid transit time, $T_{\text{mid}}$. For the fit we used a Monte-Carlo Markov chain of 350000 steps with 200 walkers and 200000 burned iterations, using the emcee Python package \citep{emcee}. Each light curve has been fitted individually and so we obtained the white light curve squared radii ratio for each transit $(R_p/R_\star)^2_{w,v}$. Initially, we fitted the white light curves using the formulae above and the uncertainties per pixel, as propagated through the data reduction process. However, it is common in HST/WFC3 data to have additional scatter that cannot be explained by the ramp model. For this reason, we scaled up the uncertainties in the individual data points, for their median to match the standard deviation of the residuals, and repeated the fitting \cite{Tsiaras2018} The resulted detrended white light curves are reported in Figure \ref{fig:white_lc} and the fit results are in Table \ref{tab:iraclis_output}. The Table shows that the $(R_p/R_\star)^2_{w,v}$ are compatible under $2 \, \sigma$ and that the standard deviation of the fitting residuals for the white light curves are significantly higher than the photon noise. The ratio between the standard deviation of the fitting residuals and the photon noise is reported in Table \ref{tab:iraclis_output} as $\bar{\sigma}$.

\begin{table*}[htb]
    \centering
    \caption{Derived parameters from the white light curves and fitting metrics: the reduced Chi-squared ($\bar{\chi^2}$), the standard deviation of the residuals with respect to the photon noise ($\bar{\sigma}$) and the auto-correlation (AC).}
    \label{tab:iraclis_output}
    \begin{tabular}{c c c c c c}
         \hline
         \hline
         Parameter &  1\textsuperscript{st} visit & 2\textsuperscript{nd} visit & 3\textsuperscript{rd} visit & 4\textsuperscript{th} visit & 5\textsuperscript{th} visit\\
        \hline
        $\left(R_p /R_\star \right)^2_{w,v}$ & $0.002430^{+0.000039}_{-0.000030}$ & $0.002430^{+0.000024}_{-0.000017}$ & $0.002440^{+0.000040}_{-0.000030}$ & $0.002632^{+0.000011}_{-0.000017}$ & $0.002352^{+0.000029}_{-0.000029}$\\
        $T_{\text{mid}} \; [BJD -2450000]$ & $7862.19152 ^{+0.00008}_{-0.00009}$ & $8020.19899 ^{+0.00004}_{-0.00004}$ & $8077.21035 ^{+0.00008}_{-0.00007}$ & $8080.46887 ^{+0.00007}_{-0.00007}$ & $8083.72719 ^{+0.00008}_{-0.00005}$ \\
        $\bar{\chi}$ & 1.18 & 1.16 & 1.17 & 1.20 & 1.17\\
        $\bar{\sigma}$ & 1.43 & 1.03 & 1.50 & 2.42 & 1.45 \\
        AC & 0.13 & 0.32 & 0.15 & 0.11 & 0.22\\
        \hline
    \end{tabular}
\end{table*}

\subsubsection{Spectral light-curves fitting}
\label{sec:spectral_light_curves}

To correct for the systematics present in the spectral light-curves of each visit, $F_{\lambda}^v(t)$, we fit each curve with a model that includes the white light curve associated, $F_{w}^v(t)$: 
\begin{equation}
   n^{\text{scan}}_\lambda \left[ 1-r_a (t-T_0) \right] \frac{F_{\lambda}^v(t)}{F_{w}^v(t)}
\end{equation}

where  $r_a$ is the coefficient of a wavelength-dependent linear slope along each HST visit and $n^{\text{scan}}_\lambda$ is the normalisation factor, that changes to $n^{\text{for}}_\lambda$, when the scanning direction is upwards, and to $n^{\text{rev}}_\lambda$ when it is downwards. In the spectral light-curve fitting, the only free parameter is R$_{\rm {p}}$/R$_\star$, while the other parameters are the same as we used for the white light-curve fitting. For the fit we used a Monte-Carlo Markov chain of 150000 steps with 100 walkers and 100000 burned iterations, using again the emcee package \citep{emcee}. 

To check the quality of our fits, we use the auto-correlation of the residuals for each light curve using the numpy correlate package. To determine a `good' value of the auto-correlation, we generated a thousand instances of random Gaussian noise and computed the auto-correlation. For the number of data points in our light curves ($\sim$70), 85\% of the time the auto-correlation of Gaussian noise is below 0.35. For each of our spectra, the auto-correlation is smaller than 0.32 (see Table \ref{tab:iraclis_output}) and thus any correlations found are consistent with those found in Gaussian noise.

%can be assumed to be no more correlated than gaussian noise. Given that by simulating data-points affected by Gaussian noise only we obtain an auto-correlation $<0.36$, our results suggest that 

We also check the success of our fit by computing the reduced Chi-squared from the comparison between the data and the model ($\bar{\chi}$) as well as the standard deviation of the residuals with respect to the photon noise ($\bar{\sigma}$). The reduced Chi-square between the spectral light curves for each visit is between 1.16 (second visit) and 1.2 (fourth visit). The averaged standard deviation of the residuals with respect to the formal photon noise is between 1.03 (second visit) and 2.42 (fourth visit), and therefore the resulted post-processing total noise is between $6 \%$ and $142 \%$ greater than the photon noise. 

\begin{figure}
    \centering
    \includegraphics[width = \columnwidth]{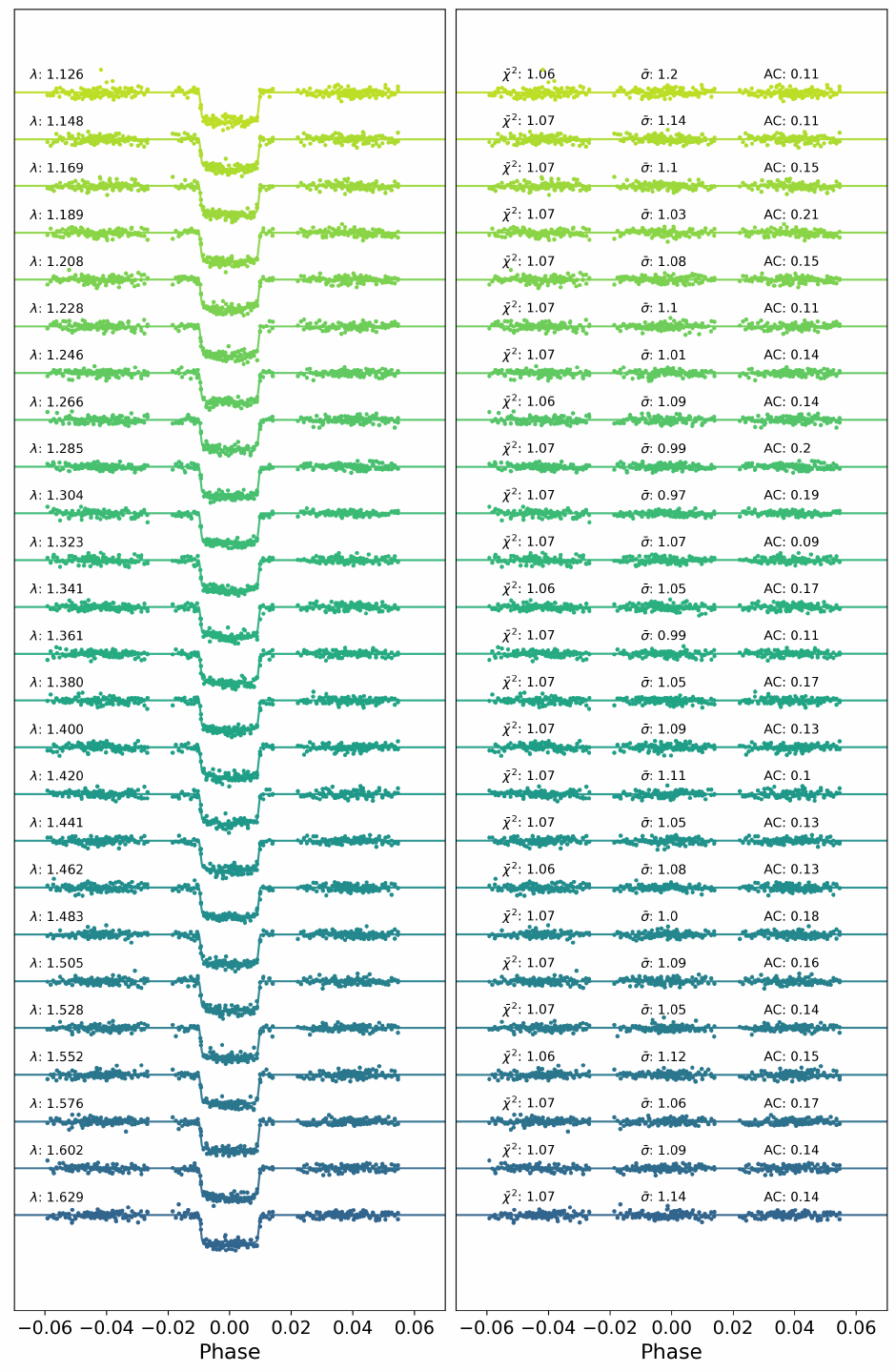}
    \caption{Spectral light curves fitted with Iraclis for the transmission spectra where, for clarity, an offset has been applied. Left: the detrended spectral light curves with best-fit model plotted. Right: residuals from the fitting with mean values for the reduced Chi-squared ($\bar{\chi^2}$), the standard deviation of the residuals with respect to the photon noise ($\bar{\sigma}$) and the auto-correlation (AC) across the five transits.}
    \label{fig:spec_lc}
\end{figure}

\begin{figure}
    \centering
    \includegraphics[width = \columnwidth]{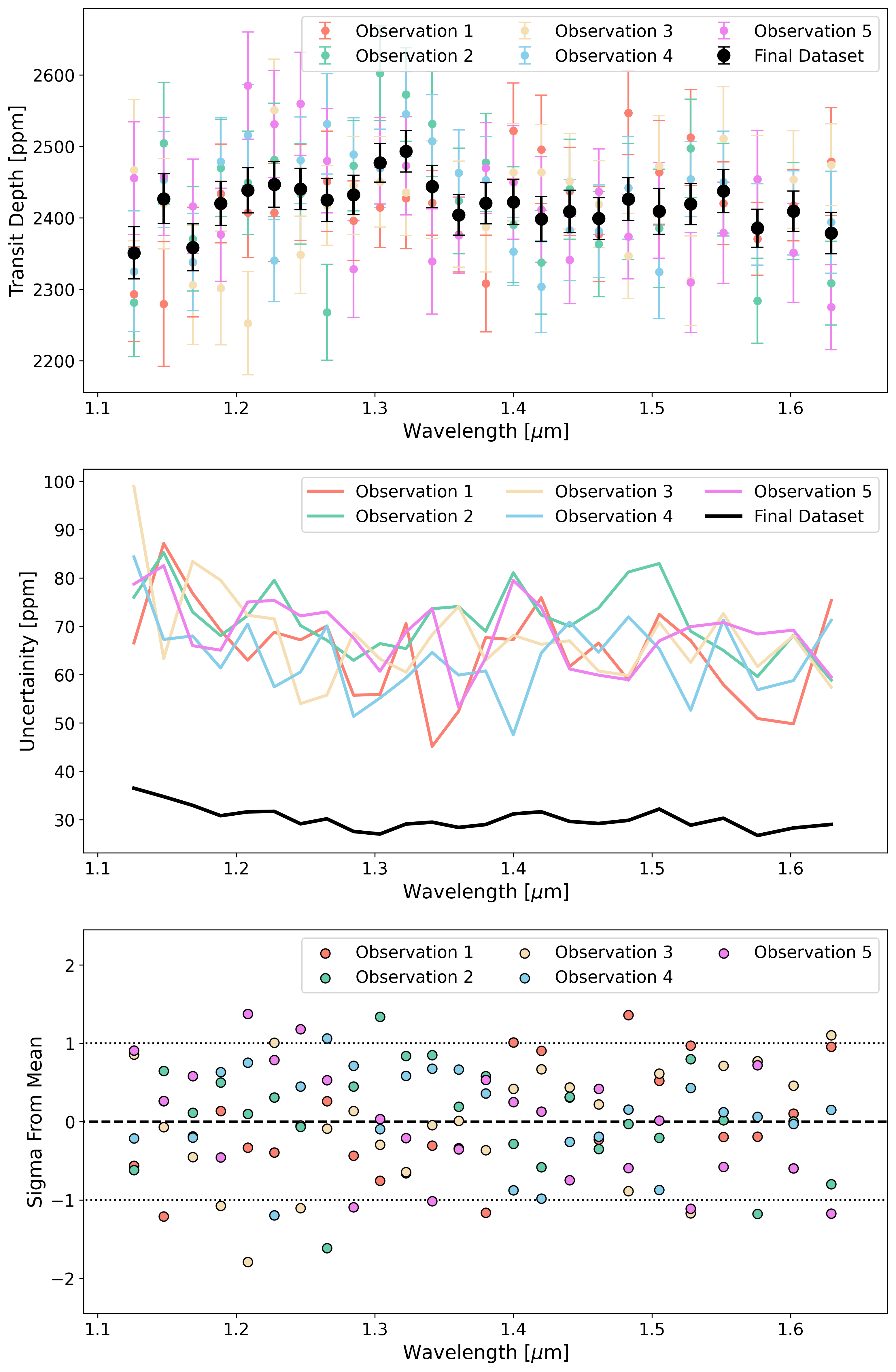}
    \caption{In the top panel are reported the transmission spectral data collected in each visit with their uncertainties (coloured data points) and the average transmission spectrum (black) obtained from their combination. The middle panel shows the uncertainty of each data point while the bottom panel shows the divergence from the mean, in sigma, of each data point.}
    \label{fig:spectrum}
\end{figure}

Then, to compare light curves obtained from different visits, $F_{\lambda}^v(t)$, we correct for offsets by subtracting each spectrum by the corresponding white light curve depth, $(R_p/R_\star)^2_{w,v}$, and adding the weighted average transit depth of all white light-curves, $(R_p/R_\star)^2_w$. Finally, we compute the weighted average from all the transit observations, $(R_p/R_\star)^2_\lambda$, reported in Table \ref{tab:iraclis_spectrum}. The spectral light curve fits are shown in Figure \ref{fig:spec_lc} while the spectrum obtained from each visit, and the final average spectrum, are shown in Figure \ref{fig:spectrum}.

We notice that in each spectral bin, the measurements in each observation are generally compatible within $1\sigma$ of the mean, as shown in Figure \ref{fig:spectrum}. Where data points are not within $1\sigma$, there are no obvious trends with the observation number or wavelength. To assess if the five transmission spectra obtained are statistically consistent, we perform a Kolmogorov-Smirnov test. We perform this test by comparing the transmission spectra two at a time, in every possible combination, to test the null hypothesis that they come from the same distribution. We use the SciPy package \citep{scipy} for the computation. We conclude that we cannot reject the null hypothesis for any of the couples because the minimum resulting p-value is $28\%$.

\begin{table}[]
    \centering
    \caption{This table reports for every spectral bin the averaged transmission spectra measured with combined the spectral light curves obtained with Iraclis, $(R_p/R_\star)^2_{\lambda}$ .}
    \label{tab:iraclis_spectrum}
    \begin{tabular}{c c c c}
         \hline
         \hline
         $ \lambda \; [\mu m]$ & $(R_p/R_\star)^2_{\lambda}$ & Uncertainty & $ \Delta \lambda \; [\mu m]$ \\
         \hline
1.1263 & 0.002325 & 0.000034 & 0.0219 \\
1.1478 & 0.002402 & 0.000033 & 0.0211 \\
1.1686 & 0.002376 & 0.000035 & 0.0206 \\
1.1888 & 0.002412 & 0.000030 & 0.0198 \\
1.2084 & 0.002453 & 0.000031 & 0.0193 \\
1.2275 & 0.002441 & 0.000031 & 0.0190 \\
1.2465 & 0.002418 & 0.000030 & 0.0189 \\
1.2655 & 0.002428 & 0.000031 & 0.0192 \\
1.2848 & 0.002425 & 0.000028 & 0.0193 \\
1.3038 & 0.002481 & 0.000028 & 0.0188 \\
1.3226 & 0.002478 & 0.000028 & 0.0188 \\
1.3415 & 0.002428 & 0.000030 & 0.0189 \\
1.3605 & 0.002408 & 0.000028 & 0.0192 \\
1.3801 & 0.002444 & 0.000029 & 0.0199 \\
1.4000 & 0.002439 & 0.000030 & 0.0200 \\
1.4202 & 0.002398 & 0.000031 & 0.0203 \\
1.4406 & 0.002388 & 0.000029 & 0.0206 \\
1.4615 & 0.002377 & 0.000028 & 0.0212 \\
1.4831 & 0.002457 & 0.000029 & 0.0220 \\
1.5053 & 0.002404 & 0.000029 & 0.0224 \\
1.5280 & 0.002438 & 0.000029 & 0.0230 \\
1.5516 & 0.002435 & 0.000030 & 0.0241 \\
1.5763 & 0.002388 & 0.000027 & 0.0253 \\
1.6021 & 0.002416 & 0.000027 & 0.0264 \\
1.6295 & 0.002418 & 0.000028 & 0.0283 \\
         \hline
    \end{tabular}

\end{table}

\subsection{Atmospheric characterisation}
\label{sec:atm_char}

To characterise the planetary atmosphere we use the retrieval code TauREx3\footnote{\url{https://github.com/ucl-exoplanets/TauREx_public}} \citep{Al-Refaie2020}, the new version of TauREx \citep{Waldmann2015b,Waldmann2015a}. The code maps the atmospheric forward model parameter space to find the best fit to the observed spectra. TauREx allow us to identify absorbers in the spectrum using line-lists from ExoMol \citep{ExoMol}, HITEMP \citep{HITEMP} and HITRAN \citep{HITRAN}. To perform the retrieval we use Multinest algorithm \citep{Feroz2011,buchner_multinest} to sample the parameter space through 1500 live points and we set the algorithm evidence tolerance to 0.5.

\subsubsection{Temperature-pressure profile }
We simulate the planetary atmosphere assuming an isothermal temperature-pressure profile with constant molecular abundances as a function of altitude. This assumption is driven by the narrow wavelength range investigated in the data, that results in restricted probed range of the planetary temperature–pressure profile \citep{Rocchetto2016}.
We calculated the equilibrium temperatures as:
 \begin{equation}
     T_{\text{eq}}= T_{\star} \left( \frac{ R_{\star}}{2\,a} \right)^{1/2} (1-A)^{1/4}
 \end{equation}
where $R_{\star}$ is the stellar radius, $a$ is the semi-major axis, $A$ is the Bond albedo. To keep in line with other studies, we adopt a standard Bond albedo of $A=0.3$ \citep{Bonfils2018}, although we are aware that the albedo is highly sensitive to the planetary properties and could therefore vary significantly \citep[e.g.][]{Marley1999,Modirrousta2021}. For the equilibrium temperature we use $T_{\text{eq}} = 520 \pm 44 \; K$, where the uncertainty comes from the error propagation on the stellar parameters and the planet’s semi-major axis from Table \ref{tab:iraclis_input}. Then for the atmospheric retrievals we use a wide range of temperature priors from $0.5 \;  T_{\text{eq}}$ to $1.5  \; T_{\text{eq}}$ (i.e. 260 - 780 K).

\subsubsection{Atmosphere composition }

For the atmosphere we use the plane-parallel approximation, building 100 plane-parallel layers to uniformly sample in log-space the pressure range from $10^{-4}$ to $10^6 \; \rm{Pa}$. We assumed a primary atmosphere of He and H$_2$ with a fixed ratio between the two molecules of 0.17, then we introduce the trace gases: H$_2$O \citep{Barton2017,Polyansky2018}, CH$_4$ \citep{Hill2013,Yurchenko2014}, CO \citep{Li2015}, CO$_2$ \citep{Rothman2010}, HCN \citep{ExoMol_HCN} and NH$_3$ \citep{ExoMol_NH3}. To perform the atmospheric fit we use as boundaries for each molecule $10^{-12}$ and $10^{-1}$ in volume mixing ratios (log-uniform prior). To fit the planet atmosphere we used two models: one with all the molecules listed above molecules and a second which also included N$_2$ \citep{western_n2}. In fact N$_2$ is a largely inactive gas over the spectral range considered and its only contribution is to the atmospheric mean molecular weight. Using such distinction between the two models we are considering a light, primary atmosphere while also exploring the potential for heavy, secondary atmosphere. We note that, as the abundances of all molecules was allowed to extend to 1 (i.e. 100\%), the retrieval without N$_2$ is also capable of resulting in a high mean molecular weight atmosphere.

Additionally, we include in all our models Rayleigh scattering and collision induced absorption of H$_2$–H$_2$ \citep{Abel2011,Fletcher2018} and H$_2$–He \citep{Abel2012}. Clouds are modeled assuming a grey opacity model and cloud top pressure bounds are set between $10^{-2}$ and $10^6$~Pa. We also set a large range of priors for the planetary radius, from $0.5 \; R_p$ to $1.4 \; R_p$ referring to the literature value reported in Table \ref{tab:iraclis_input}.  The planetary radius is assumed to be equivalent at $10^6 \; $Pa pressure.

To assign a significance to our detection, we use the Bayes Factor between the nominal atmospheric model and a model which contains no active trace gases, Rayleigh scattering or collision induced absorption. We perform a retrieval where no molecular absorbers are active which provides a flat line model to test the significance of retrievals including molecular opacities.

%This value is computed as the difference of the two model logarithmic evidences and is then translated into a statistical $\sigma$ significance by using Table 2 of \citet{Benneke2013}. 

\section{Results}

\label{sec:atm_ret}

The final spectrum recovered is extremely flat, with few deviations from a flat line. Nevertheless, we conducted retrievals to explore the possibility that any of these minor features could be attributed to molecular species. While formation and evolution theories suggest a hydrogen dominated atmosphere is unlikely for this planet, we started by exploring this possibility. Our initial retrieval, conducted with a hydrogen-dominated atmosphere containing clouds and the molecules discussed in Section \ref{sec:atm_char}, found no evidence of features. The retrieval including N$_2$ came to similar results, with the spectrum essentially being fitted by a grey cloud deck alone in both cases. The Bayesian evidence for each retrieval, log(E) = 218.44 and log(E) = 218.38 for retrievals with and without N$_2$ respectively, shows that the cloud only model (log(E) = 218.52) provides the best fit to the data.

\begin{figure}
    \centering
    \includegraphics[width=\columnwidth]{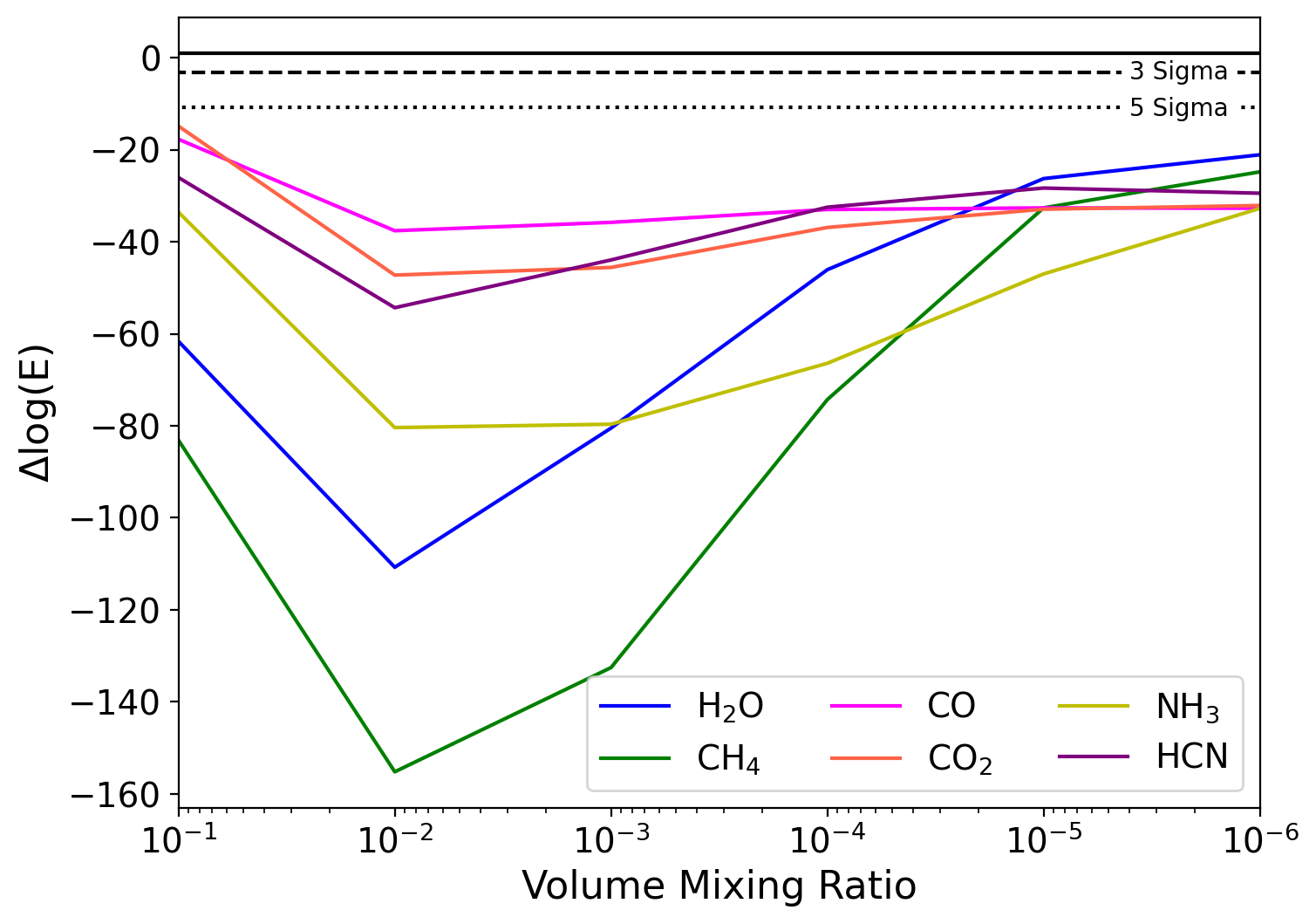}
    \caption{Comparison of the log evidence for a cloudy atmosphere to that of single molecule retrievals where the abundance of said molecule was fixed and no clouds were included. In each case, the cloudy model is preferred to $>$5$\sigma$.}
    \label{fig:log_evidence}
\end{figure}

\begin{figure}
    \centering
    \includegraphics[width=\columnwidth]{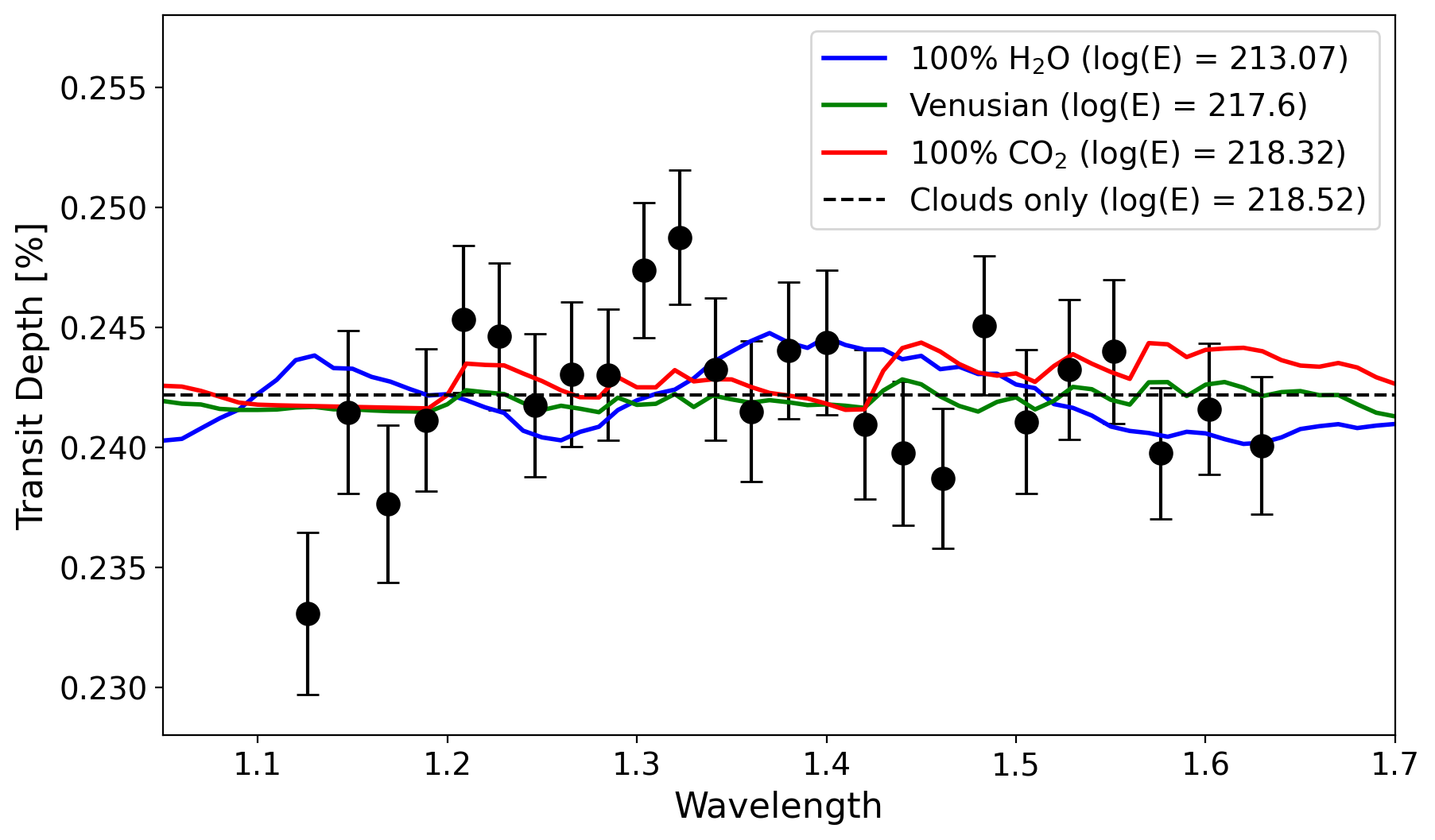}
    \caption{Best-fit spectra for the secondary atmosphere models. The cloud only model (i.e. a flat line) still provides the best-fit to the spectrum obtained.}
    \label{fig:secondary_spec}
\end{figure}

\begin{figure*}
    \centering
    \includegraphics[width = .8\textwidth]{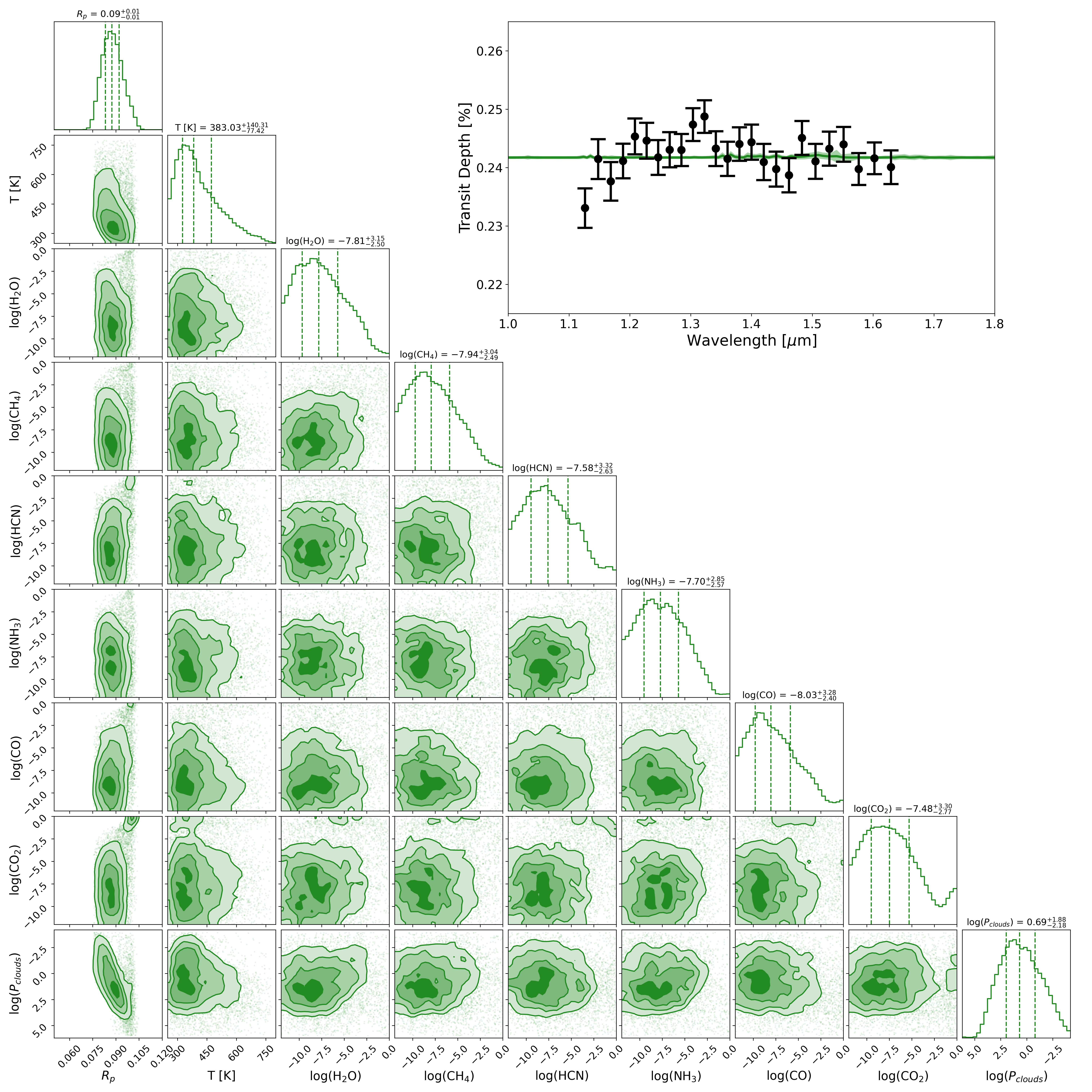}
    \caption{Inset top right: reported the transmission spectra extracted with Iraclis software from the 5 HST visits of GJ\,1132\,b. In black is reported the average transmission spectra while the green line is the spectrum retrieved with TauREx and the shaded region represents the $3\sigma$ confidence level. Left: the retrieval posteriors for GJ\,1132\,b which show no evidence for molecular species.}
    \label{fig:taurex_retrieval}
\end{figure*}

To better show the noncompliance of our data with molecular features, we performed retrievals for clear hydrogen-dominated atmospheres with only one molecule which was forced to mixing ratios from $10^{-6}$ to $10^{-1}$. The latter is set as a rough boundary between primary and secondary atmospheres (i.e. a mean molecular weight $>>$ 2.3). We use the Bayesian evidence from these retrievals, given in Figure \ref{fig:log_evidence}, to rule them out to a given significance. We ran these for H$_2$O, NH$_3$, CO, CO$_2$, CH$_4$ and HCN and, in every case, the cloud only model provided a better fit to the data to $>$5$\sigma$. Hence, our results suggest we can rule out a clear, primary atmosphere with high confidence: if GJ\,1132\,b hosts a primary atmosphere (i.e. one dominated by hydrogen and helium), according to our spectrum, the planet must be completely overcast.

We also attempted to fit several models with secondary atmospheric compositions. These were atmospheres composed entirely of H$2$O and CO$_2$ (i.e. VMR$_{\rm H_2O}$ = 1 and VMR$_{\rm CO_2}$ = 1), as well as an atmosphere similar to that of Venus (with volume mixing ratios of VMR$_{\rm CO_2}$ = 0.965, VMR$_{\rm H_2O}$ = 2e-5, VMR$_{\rm CO}$ = 1.7e-5, VMR$_{\rm SO_2}$ = 1.5e-4). For each of these, a cloud-free atmosphere was assumed and the molecular abundances were again fixed. Thus, the only free parameters were the planet radius and the temperature. The best-fit spectrum in each case is given in Figure \ref{fig:secondary_spec} although with the Bayesian evidence for each setup. Again, the best fitting model is that of a flat line, with a cloud free, 100\% H$_2$O atmosphere being ruled out to $>$3 sigma. However, while providing worse fits to the data than the cloud-only model, the clear, 100\% CO$_2$ and Venusian atmosphere models cannot be definitively ruled out with the current dataset.

Hence, no evidence for molecular features could be extracted from this dataset. For completeness we include the posterior distribution for our baseline retrieval in Figure \ref{fig:taurex_retrieval}. Our results suggest that the atmosphere of GJ\,1132\,b is likely to be cloudy but that certain enriched atmospheres, with small scale-heights that led to only minor features over the HST WFC3 range, could also explain the data. A final possibility, which is compatible with our spectrum, is that the planet hosts an atmosphere too thin to be detected and that we are measuring the transit depth caused by the planet's solid surface.

%\begin{table}
%    \centering
%    \caption{Log evidence comparison for the fitted atmosphere models. As the cloud only model has the highest log evidence, it is preferred to the other models.}
 %   \label{tab:retrieval_comparison}
 %   \resizebox{\columnwidth}{!}{
 %   \begin{tabular}{lc}
%    \hline \hline
 %       Atmospheric Model & Log Evidence\\
 %       \hline \hline
%        clouds only & 216.76 \\
%        H$_2$O,NH$_3$,CO,CO$_2$ \& CH$_4$ & 215.97 \\
%        H$_2$O,NH$_3$,CO,CO$_2$, CH$_4$ \& N$_2$ & 216.12 \\
%        Venusian & 215.70 \\
    
%%\hline

%\hline
%    \end{tabular}
%    }
%\end{table}

%\begin{table}
%\centering
%\caption{Log evidence comparison for clear hydrogen/helium dominated atmospheres with fixed molecular abundances against the flat line model.}
%\label{tab:primary_atmo}
%\begin{tabular}{lccc}\hline\hline
%\multirow{2}{*}{H$_{\rm 2}$O} & -3 & 168.65 & $>$5 \\
%                            & -6 & 200.83 & $>$5 \\\hline
                            
%\multirow{2}{*}{NH$_{\rm 3}$} & -3 & 141.64 & $>$5 \\
%                            & -6 & 188.64 & $>$5 \\\hline
                            
%\multirow{2}{*}{CH$_{\rm 4}$} & -3 & 94.3 & $>$5 \\
%                            & -6 & 196.61 & $>$5 \\\hline
                            
%\multirow{2}{*}{CO} & -3 & 187.58 & $>$5 \\
%                      & -6 & 191.10 & $>$5 \\\hline
                      
%\multirow{2}{*}{CO$_{\rm 2}$} & -3 & 191.93 & $>$5 \\
%                            & -6 & 191.27& $>$5 \\\hline

%\multicolumn{2}{c}{Cloud Only Model:} & 216.76 & -\\
%\hline\hline
%\end{tabular}
%\end{table}

\section{Discussion}

\subsection{Statistical goodness of fit analysis}

Due to the absence of molecular features in the measured spectrum, we statistically compare the data reported in Table \ref{tab:iraclis_spectrum} with a constant value defined as the weighted average of the planetary-to-stellar radii ratios. By using chi squared statistics, we obtain $\chi^2 = 26$ with $\nu = 25-1 = 24$ degrees of freedom. The obtained reduced chi squared value is $\chi_{\nu}^2 = 26/24 = 1.08$, with an uncertainty of $\sigma_{\chi_{\nu}^2} = 0.29$. Because $\chi_{\nu}^2 \sim 1$, we cannot exclude the possibility that the measured data comes from a constant distribution as the flat line is a good model to describe the data. Therefore, the collected data does indeed show no detectable molecular features. 

Because we compare different observations to obtain the spectrum in Table \ref{tab:iraclis_spectrum}, we perform a Jack-knife analysis \citep{quenouille1949,quenouille1956,Tukey1958}. This consists of taking the original dataset, $X$, which is composed of $n$ samples, and creating $n$ new datasets, $x_{i}$. Each new dataset is like the original except that it has $n-1$ samples. The removed sample is different for each new dataset so that no two are identical. In our case we have 5 observations so $n=5$. For each of these new samples we compute the average spectrum, as described in sec \ref{sec:spectral_light_curves}, and then we compare it with its weighted average value, $\hat{x_i}$. In an identical manner as before, we make use of the reduced chi squared value $\chi_{\nu}^2$ as a metric for the goodness of the fit. The results are reported in Tab. \ref{tab:jk}. All of our $\chi_{\nu}^2$ are $\sim 1$ indicating a very good fit. This confirms our results and demonstrates that we are not biased in our analysis. 

\begin{table}
    \centering
    \caption{Reduced chi squared ($\chi_{\nu}^2$) from Jack-knife analysis. The first column reports the observation that has been remove to the data set before computing the chi squared. The second column reports the resulted reduced chi squared of the comparison between the dataset average spectrum and its weighted average value, for the data set without the removed observation.}
    \label{tab:jk}
    \begin{tabular}{cc}\hline\hline
    Rem. obs. & $\chi_{\nu}^2$ \\ \hline\hline
         1 & 1.3 \\ 
         2 & 0.6 \\ 
         3 & 1.6 \\ 
         4 & 1.0 \\ 
         5 & 1.2 \\ 
         \hline\hline
    \end{tabular}
\end{table}

We also computed the Jack-knife bias as  $bias = (n-1)(\hat{X}-\hat{x})= 3e-7$, where $\hat{X}$ is the average radii ratio computed from the $X$ sample, and $\hat{x}$ is the average of the mean radii ratios $\hat{x_i}$. The measured bias is two orders of magnitude smaller than the uncertainties and can therefore be neglected.

\subsection{The Physical Implications of our Results}

As explained previously, from mass and radius measurements it is not always possible to accurately determine the bulk composition of an exoplanet. However, there are a few clues that could help us evaluate whether GJ~1132~b could have a hydrogen atmosphere or not. Specifically, one could consider the effects of X-ray and ultraviolet (XUV) irradiation from the host star and its effects on a primordial atmosphere. Performing a backwards reconstruction of the maximum amount of hydrogen that could be lost by XUV irradiation is beyond the scope of this study, but it is worth discussing. 

Consider for example the super-Earth GJ~357~b. Having a mass, radius, and temperature of $\rm \simeq 1.84 M_{\oplus}$, $\rm \simeq 1.22 R_{\oplus}$, and $\rm \simeq 500~ K$ \citep{Luque2019}, respectively, its properties are relatively similar to those of GJ~1132~b. However, what makes GJ~357~b unique is that it orbits a very low active M-type star \citep{Modirrousta2020c}. In spite of the abnormally low XUV levels, a careful backwards evaporation reconstruction model shows that up to $\sim$ 38 M$_{\oplus}$ of hydrogen could have been lost \citep{Modirrousta2020c}. Although GJ~357~b was probably born with a hydrogen envelope significantly smaller than this (perhaps M$_{atm} \lesssim 0.02~M_{\oplus}$ \citet{Ikoma2012,Chachan2018}), this calculation shows that even stars with very low activity levels could completely strip off the primordial atmosphere of a planet. While the activity level of GJ~1132 is not known, statistically speaking it is most probably higher than that of GJ~357~b \citep[e.g.][]{Penz2008(1),Sanz-Forcada2011}, so by comparison one can infer that GJ~1132~b most probably lost its hydrogen envelope. Of course, other effects such as magnetism \citep[e.g.][]{Matsakos2015}, magma-atmosphere exchanges \citep[e.g.][]{Chachan2018}, and migration \citep[e.g.][]{Nayakshin2012} may have lowered the evaporation rates, but large mass-losses would be expected nonetheless. 
Therefore, based on the mass-radius of GJ~1132~b, our current understanding of XUV-induced evaporation, and our spectroscopic results, a strong argument can be made that GJ~1132~b is a telluric body lacking a primordial envelope that instead might host a secondary atmosphere. However, we do acknowledge that other setups are plausible. For instance, GJ~1132~b could be an airless rocky planet \citep[e.g.][]{Modirrousta2021} or a rocky planet without clouds but an atmosphere that is too thin to be detected in these data. Notwithstanding, these configurations may be less probable given that geological outgassing is expected to generate thick secondary atmospheres with clouds \citep{Kite2009,Noack2017,Dorn2018}.

\subsection{Comparison with previous works}

\begin{figure}
    \centering
    \includegraphics[width = \columnwidth]{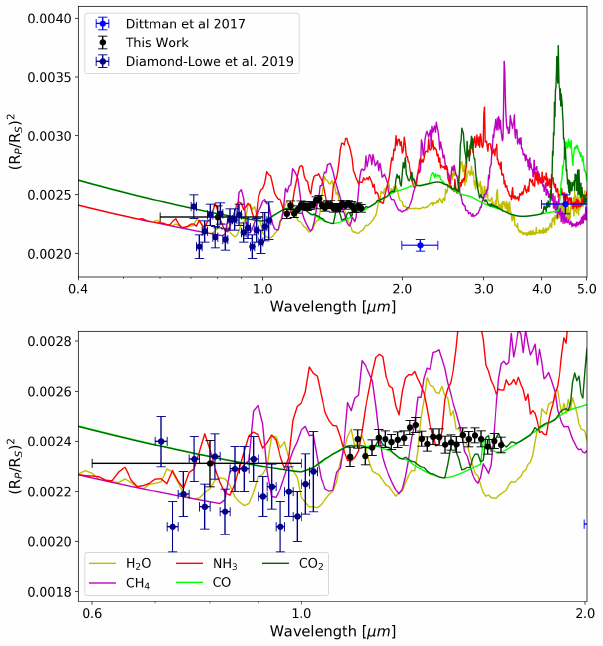}
    \caption{Transit depths from various studies on GJ\,1132\,b, as listed in Table \ref{tab:radii_comparison}, and the spectrum recovered in this work. Also plotted are forward models which assume a clear H/He atmosphere with various molecules, each at an abundance of volume mixing ratios of $10^{-3}$. We note that we do not directly compare the models to the combined dataset due to the possibility of incompatibilities between them.}
    \label{fig:radii_comparison_spectra}
\end{figure}

To compare our measurements with previous works we adopt the radii ratios published in recent papers. We use the measurements from Table 3 of \citet{Dittmann2017} %, Figure 9 of \citet{Southworth2017} 
and Table 6 of \citet{Diamond-Lowe2018}, and we compute the  $(R_p/R_\star)^2$ from them, where needed. Additionally, GJ\,1132\,b has also been studied by TESS and we used the pipeline from \cite{edwards_orbyts} to download, clean and fit the 2 minute cadence Pre-search Data Conditioning (PDC) light curves \citep{smith_pdc,stumpe_pdc1,stumpe_pdc2}. Then we report all the data in Table \ref{tab:radii_comparison} and in Figure \ref{fig:radii_comparison_spectra}. 

While it has become common to combine data from different instruments there may be an offset between the datasets. These offsets can occur due to imperfect correction of instrument systematics, from the use of different orbital parameters or limb darkening coefficients during the light curve fitting or from stellar variability or activity \citep[e.g.][]{stevenson_gj436,morello_ldc,stevenson_wasp12,Tsiaras2018,yip_lc,bruno_spots,ares3,murgas_offset,changeat_k11,yip_2021,schlawin_nircam}. As there is no wavelength overlap between our HST observations and the ground-based data we cannot be certain of the compatibility of the observations. 

The transit depth from the TESS data agrees with the data from \citet{Diamond-Lowe2018} (see Figure \ref{fig:radii_comparison_spectra}). Nevertheless, while the datasets could potentially be compatible, we err on the side of caution and do not perform a joint fit. We note that the study by \citet{Diamond-Lowe2018} also concluded that GJ\,1132\,b could not have a clear, primordial envelope and thus the studies are in agreement on this conclusion. Additionally, the Spitzer transit from \citet{Dittmann2017} exhibits a similar transit depth to ours, but again we do not include it in a joint fit. Finally, we note that \citet{Southworth2017} claimed the detection of an atmosphere due to the modulation of several ground-based photometric measurements but, given the precision obtained in \citet{Diamond-Lowe2018} was higher than in said study, we do not include it in our plots.

\begin{table}[]
    \centering
    \caption{Collection of transit depths of GJ\,1132\,b from other instruments. 
    Data from \citet{Dittmann2017} are from Table 3 and for \citet{Diamond-Lowe2018} we use data from Table 6.}
    \label{tab:radii_comparison}
    \begin{tabular}{l c c c}
    \hline
    \hline
    Instrument & $\lambda \; [\mu m]$ & $(R_p/R_\star)^2$ & Reference \\ 
    \hline
Spitzer & 4.50 & $0.00242 \pm 0.00008$ & D17 \\
MEarth & 2.19 & $0.00207 \pm 0.00005$ & D17 \\
LDSS3C & 0.71 & $0.00240 \pm 0.00010$ & DL18 \\
LDSS3C & 0.73 & $0.00206 \pm 0.00010$ & DL18 \\
LDSS3C & 0.75 & $0.00219 \pm 0.00009$ & DL18 \\
LDSS3C & 0.77 & $0.00233 \pm 0.00009$ & DL18 \\
LDSS3C & 0.79 & $0.00214 \pm 0.00009$ & DL18 \\
LDSS3C & 0.81 & $0.00234 \pm 0.00009$ & DL18 \\
LDSS3C & 0.83 & $0.00212 \pm 0.00009$ & DL18 \\
LDSS3C & 0.85 & $0.00229 \pm 0.00009$ & DL18 \\
LDSS3C & 0.87 & $0.00229 \pm 0.00009$ & DL18 \\
LDSS3C & 0.89 & $0.00233 \pm 0.00009$ & DL18 \\
LDSS3C & 0.91 & $0.00218 \pm 0.00008$ & DL18 \\
LDSS3C & 0.93 & $0.00222 \pm 0.00009$ & DL18 \\
LDSS3C & 0.95 & $0.00206 \pm 0.00010$ & DL18 \\
LDSS3C & 0.97 & $0.00220 \pm 0.00010$ & DL18 \\
LDSS3C & 0.99 & $0.00210 \pm 0.00010$ & DL18 \\
LDSS3C & 1.01 & $0.00223 \pm 0.00012$ & DL18 \\
LDSS3C & 1.03 & $0.00228 \pm 0.00016$ & DL18 \\
TESS & 0.8 & $0.002312 \pm 0.000093$ & This Work\\ \hline
\multicolumn{4}{c}{D17: \citet{Dittmann2017}}\\
\multicolumn{4}{c}{DL18 : \citet{Diamond-Lowe2018}}\\
    \hline
    \end{tabular}
\end{table}

During the review process of this paper, an independent study also analysed the same HST WFC3 data of GJ\,1132\,b. \cite{Swain2021} found evidence for a slope over this wavelength region, attributed to an H$_2$ dominated atmosphere with hazes, as well as spectral features that were proposed to be due to absorption by CH$_4$ and HCN. Their work suggested GJ\,1132\,b had lost its primordial envelope and gained a second atmosphere through volcanic processes that released H$_2$ captured in the early age.

As our extracted spectrum differs greatly from that of \cite{Swain2021}, we conducted an independent analysis of the datasets with an additional open-source pipeline: the Calibration of trAnsit Spectroscopy using CAusal Data (CASCADe\footnote{\url{ https://jbouwman.gitlab.io/CASCADe/}}). While Iraclis has been developed specifically for analysing HST WFC3, CASCADe is an instrument independent reduction pipeline and has been applied to both Hubble and Spitzer datasets. The CASCADe pipeline starts the data reduction with the ima intermediate data product, which were produced by the CALWFC3 data reduction pipeline (note that Iralcis takes the raw data and applies calibration steps itself). CASCADe implements a novel data driven method, pioneered by \citet{scholkopf_2016}, utilising the causal connections within a data set to calibrate the spectral time-series data. For a full description of the pipeline steps, we refer the reader to \citep{carone_wasp117}.

\begin{figure}
    \centering
    \includegraphics[width=\columnwidth]{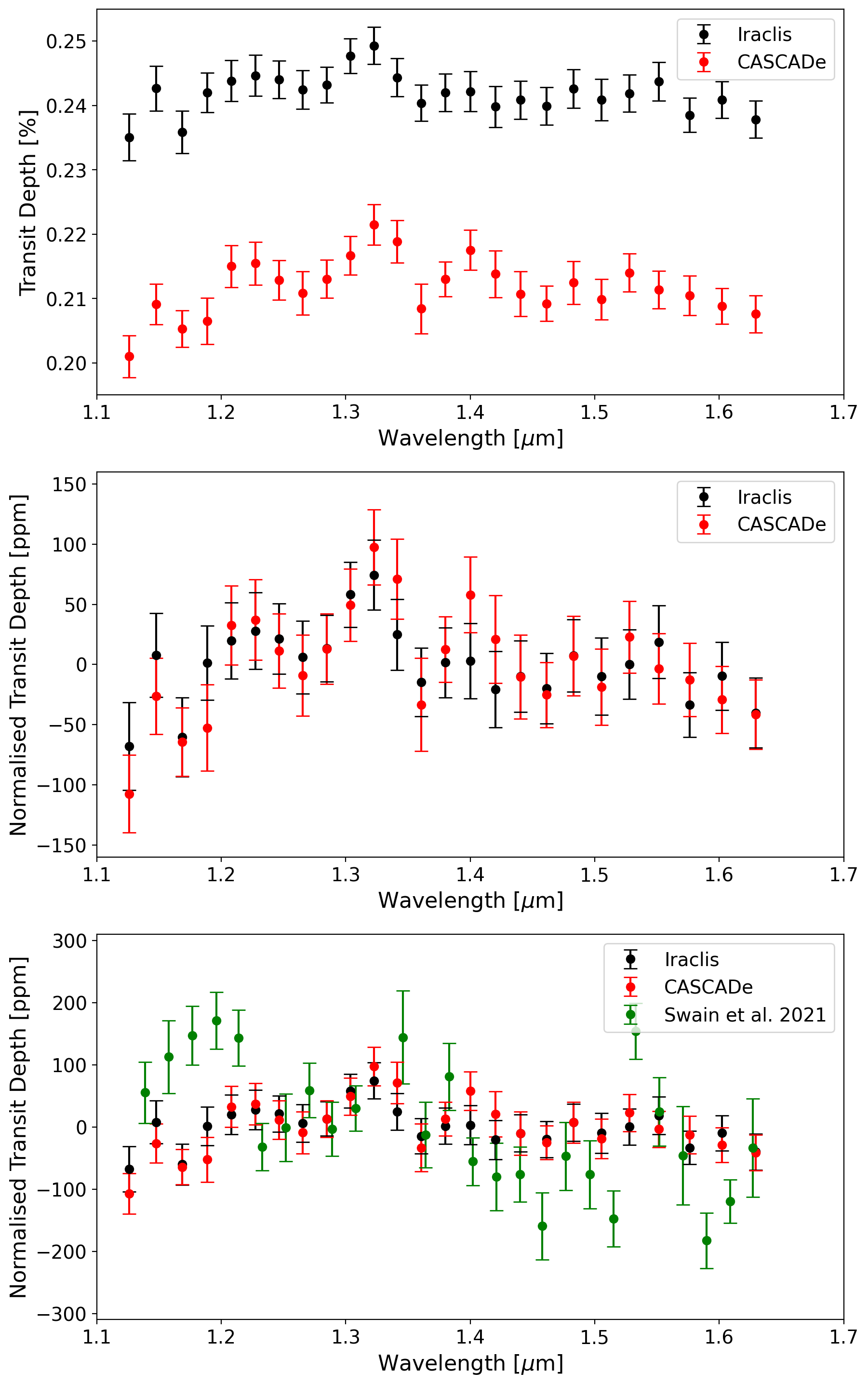}
    \caption{Comparison of the results from the data reduction and spectrum extraction undertaken here using Iraclis and CASCADe. While there is an offset between the spectra, normalising them by their mean transit depth shows they are consistent to within 1$\sigma$. However, the spectra recovered with both pipelines here differs significantly to that from \citet{Swain2021}.}
    \label{fig:swain_comp}
\end{figure}

We ran CASCADe using the same planet parameters and limb darkening coefficients as discussed in Section 2.1. A comparison between these spectra, and to that obtained by \citet{Swain2021}, is given in Figure \ref{fig:swain_comp}. We immediately notice a vertical offset in the spectra obtained by Iraclis and CASCADe. The offset is likely to be caused by differences in the correction of the systematics and such offsets between different pipelines has been seen before, for example for WASP-117\,b \citep{anisman_w117,carone_wasp117} and KELT-11\,b \citep{changeat_k11,colon_k11}. The finding of this offset provides further evidence that combining instruments without wavelength overlap is dangerous \citep{yip_lc,yip_2021}

Despite this offset, we note that the spectral features in the CASCADe spectrum, or lack thereof, are highly similar to those obtained with Iraclis. By subtracting the mean from each spectrum, we show that the recovered data points are all within 1$\sigma$. Comparing these to \cite{Swain2021}, we clearly highlight the disparity between our study and theirs: whereas they uncover spectral variations of $\pm$200\,ppm from the mean transit depth, both our pipelines yield spectra where 21 of the 25 data points (84\%) lie within $\pm$50\,ppm of the mean transit depth, with no data points being more than 75\,ppm from the mean.

The exact cause of the difference in the recovered transmission spectra between this study and that of \citet{Swain2021} is hard to discern without detailed one to one comparisons of the pipelines utilised. Iraclis and CASCADe are both open-source, well-used and have been validated against other results within the literature. Additionally, their approach to the data calibration and reduction, especially in the fitting of the systematic trends that are encountered within HST WFC3 data, are utterly different and so achieving almost identical spectra with these pipelines leads us to have confidence in our results. Our team is working closely with those from \citet{Swain2021} to resolve the discrepancy between our work and theirs.

\begin{figure}
    \centering
    \includegraphics[width = \columnwidth]{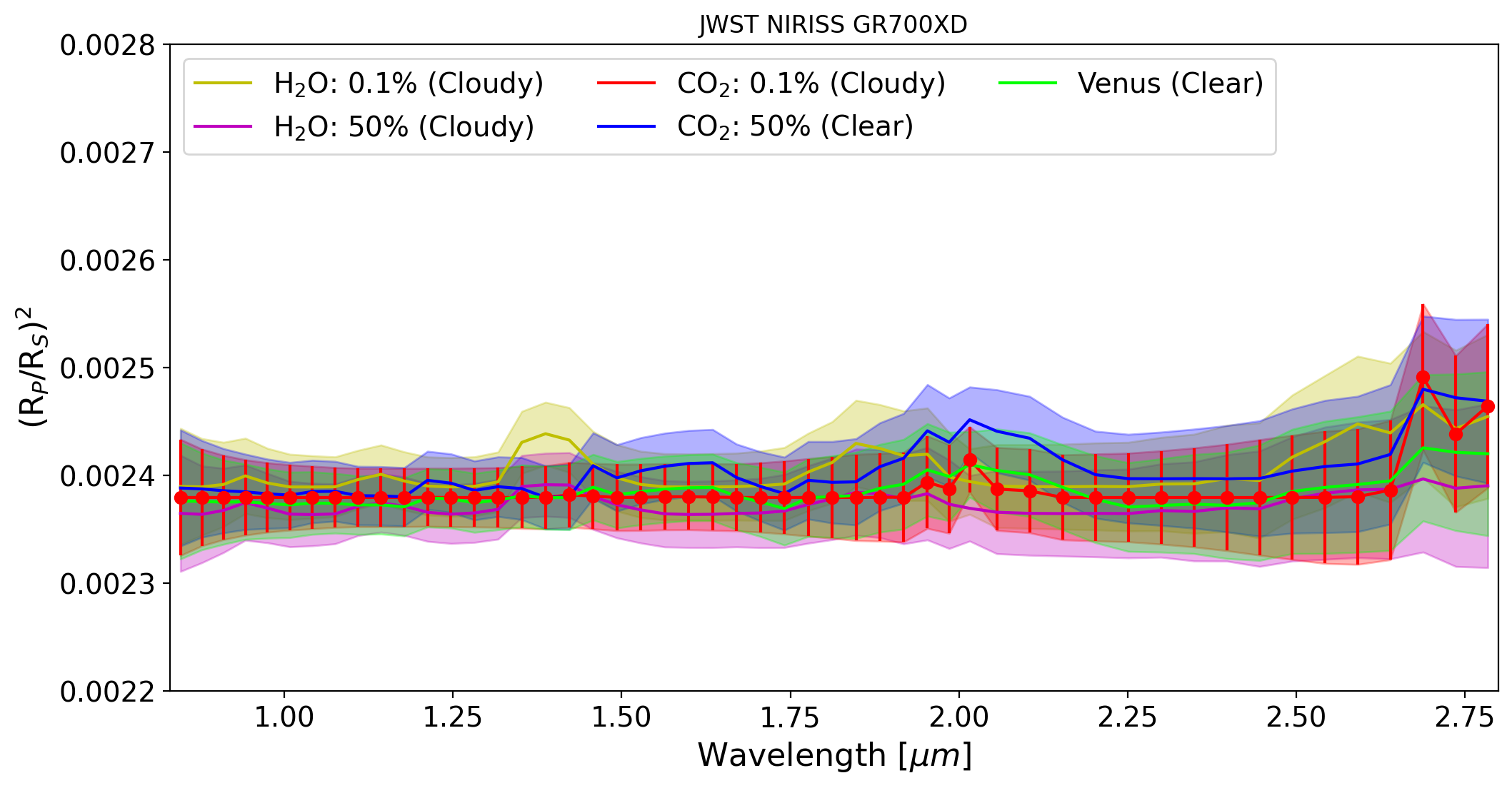}
    \includegraphics[width = \columnwidth]{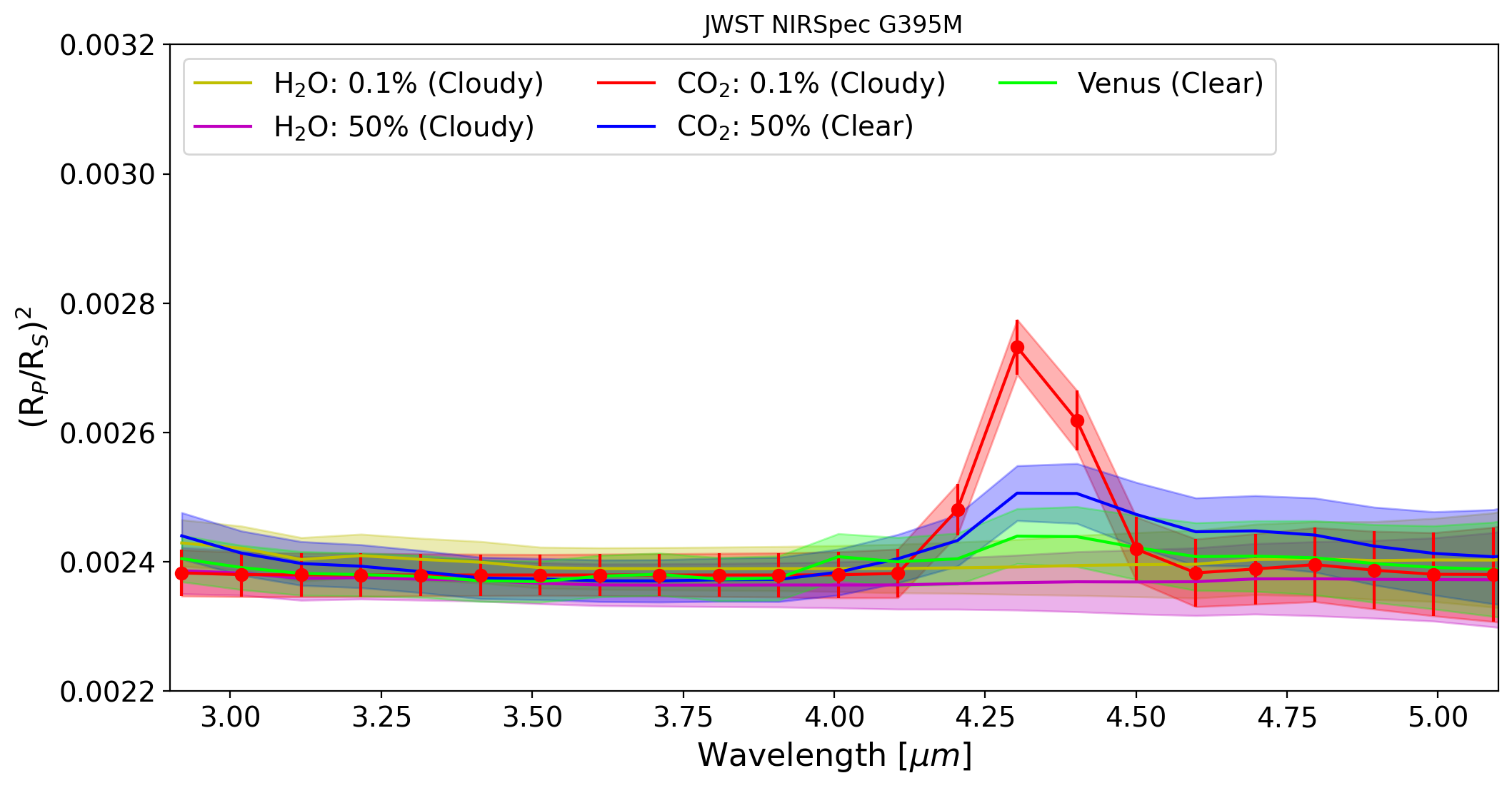}
    \caption{Simulated JWST observations for various atmospheric types which are consistent with current data. The coloured regions show the 1$\sigma$ errors on the observation which has been binned to R$\sim$50.}
    \label{fig:jwst_sim}
\end{figure}

\subsection{Future missions}

Future missions will offer increased sensitivity and wider spectral coverage. This will be key in the hunt for atmospheric features on smaller planets. While Ariel will be able to characterise H/He dominated atmospheres \citep{ariel_tl}, or rule out their presence on small worlds, it may struggle to provide additional constraints on GJ\,1132\,b given its lack of a clear H/He atmosphere and spectral features. Hence, we focus on JWST and show in Figure \ref{fig:jwst_sim} the data that could be obtained from 1 single transit with either NIRISS GR700XD or NIRSpec G395M, modelled using ExoWebb, an adapted version of the Terminus tool described in \citet{terminus} which uses the Pandeia engine \citep{pandeia}. The coloured regions show the 1$\sigma$ errors on the observation which has been binned to R$\sim$50 and, as they can rarely be distinguished, this suggests that even JWST may struggle to disentangle different atmospheric types or provide significant evidence for molecular features. With the exception of the Venus-like case, the forward models assume a H/He envelope with the addition of the stated molecule. The cloud deck for the 0.1\% H$_2$O and CO$_2$ atmospheres was set to 100 Pa (0.001 Bar), while 10 Pa (0.0001 Bar) was included for the 50\% H$_2$O case.

\section{Conclusion}

We present the data analysis of five spectroscopic observations of GJ\,1132\,b obtained with the G141 grism of the WFC3 onboard of the HST. We extracted the planetary transmission spectra with Iraclis pipeline and we attempted to retrieve the atmospheric composition using TauREx3. Our findings agree with those of \citet{Diamond-Lowe2018} and the transmission spectrum we obtain from our data shows no molecular features in the investigated wavelength range. We compared the spectrum with different atmospheric types to verify the non compliance with any molecular content at the data precision and we concluded that is compatible with a flat transmission spectrum only. Future astronomical missions, such as JWST, will help further constrain the atmospheric properties of GJ\,1132\,b although multiple observations may be required for spectral features to be discerned. While it may be difficult to understand its true nature, GJ\,1132\,b remains an interesting candidate for future atmospheric studies.

\vspace{3mm}
\textbf{Acknowledgments:}

We thank our anonymous referee for their insightful comments which have improved the quality of our work. Additionally, we thank Mark Swain and his team for open and constructive talks surrounding the analysis of these datasets.

This work was  realised as part of ``ARES Ariel School" in Biarritz in 2019. The school was organised by JPB, AT and IW with the financial support of CNES. JPB acknowledges the support of the University of Tasmania through the UTAS Foundation and the endowed Warren Chair in Astronomy, Rodolphe Cledassou, Pascale Danto and Michel Viso (CNES). WP, TZ, and AYJ have received funding from the European Research Council (ERC) under the European Union's Horizon 2020 research and innovation programme (grant agreement n$^\circ$ 679030/WHIPLASH and n$^\circ$ 758892/ExoAI). SW was supported through the STFC UCL CDT in Data Intensive Science (grant number ST/P006736/1). LVM and DMG acknowledge the financial support of the ARIEL ASI grant n. 2018-22-HH.0. BE, QC, MM, AT and IW acknowledge funding from the European Research Council (ERC) under the European Union's Horizon 2020 research and innovation programme grant ExoAI (GA No. 758892) and the STFC grants ST/P000282/1, ST/P002153/1, ST/S002634/1 and ST/T001836/1. GG acknowledges the financial support of the 2017 PhD fellowship programme of INAF. RB is a PhD fellow of the Research Foundation -- Flanders (FWO). DB acknowledges financial support from the ANR project "e-PYTHEAS" (ANR-16-CE31-0005-01). NS acknowledges the support of the IRIS-OCAV, PSL. MP acknowledges support by the European Research Council under Grant Agreement ATMO 757858 and by the CNES. OV thanks the CNRS/INSU Programme National de Plan\'etologie (PNP) and CNES for funding support. JB acknowledges support from the European Research Council under the European Union's Horizon 2020 research and innovation program ExoplANETS-A (GA No.~776403). GM has received funding from the European Union's Horizon 2020 research and innovation programme under the Marie Skłodowska-Curie grant agreement No. 895525.

\vspace{3mm}
\textbf{Data:}
This work is based upon observations with the NASA/ESA Hubble Space Telescope, obtained at the Space Telescope Science Institute (STScI) operated by AURA, Inc. The publicly available HST observations presented here were taken for proposal 14758, led by Zach Berta-Thompson \citep{berta_prop}. These were obtained from the Hubble Archive which is part of the Mikulski Archive for Space Telescopes. This paper also includes data collected by the TESS mission which is funded by the NASA Explorer Program. TESS data is also publicly available via the Mikulski Archive for Space Telescopes (MAST).

\vspace{3mm}
\textbf{Software:} Iraclis \citep{Tsiaras2016b}, TauREx3 \citep{Al-Refaie2020}, pylightcurve \citep{tsiaras_plc}, ExoTETHyS \citep{Morello2020}, Astropy \citep{astropy}, h5py \citep{hdf5_collette}, emcee \citep{emcee}, Matplotlib \citep{Hunter_matplotlib}, Multinest \citep{multinest,buchner_multinest}, Pandas \citep{mckinney_pandas}, Numpy \citep{oliphant_numpy}, SciPy \citep{scipy}.

\bibliographystyle{yahapj}
\bibliography{main}

\end{document}